\documentclass[12pt]{article}
\usepackage[hyperfootnotes=false]{hyperref}
\usepackage{epsfig}
\usepackage{amsmath}
\usepackage{amssymb}
\usepackage{caption}
\numberwithin{equation}{section}

\usepackage{setspace}
\usepackage{graphicx}
\usepackage{color}
\usepackage{authblk}

\usepackage{subcaption}
\usepackage{braket}

\def\be{\begin{equation}}
\def\ee{\end{equation}}
\def\ba#1\ea{\begin{align}#1\end{align}}
\def\bg#1\eg{\begin{gather}#1\end{gather}}
\def\bm#1\em{\begin{multline}#1\end{multline}}
\def\bmd#1\emd{\begin{multlined}#1\end{multlined}}

\def\b{\beta}

\def\pa{\partial}

\def\({\left(}
\def\){\right)}
\def\[{\left[}
\def\]{\right]}
\def\<{\langle}
\def\>{\rangle}

\def\nref#1{(\ref{#1})}

\newcommand{\beq}{\begin{equation}}
\newcommand{\eeq}{\end{equation}}
\newcommand{\beqn}{\begin{eqnarray}}
\newcommand{\eeqn}{\end{eqnarray}}
\newcommand{\cA}{\mathcal{A}}
\newcommand{\cB}{\mathcal{B}}
\newcommand{\cC}{\mathcal{C}}
\newcommand{\cf}{t}
\newcommand{\cp}{\phi}

\newcommand{\cO}{\mathcal{O}}

\title{Emergent classical spacetime from microstates of an incipient black hole}
\author[1,2]{Vijay Balasubramanian}
\author[3]{David Berenstein}
\author[4]{Aitor Lewkowycz}
\author[5]{Alexandra Miller}
\author[1]{Onkar Parrikar}
\author[2,1]{Charles Rabideau}
\vspace{1cm}
\affil[1] {\small{David Rittenhouse Laboratory, University of Pennsylvania, Philadelphia PA, 19104, U.S.A.}}
\affil[2]{\small{Theoretische Natuurkunde, Vrije Universiteit Brussel (VUB), and
International Solvay Institutes, Pleinlaan 2, B-1050 Brussels, Belgium.}}
\affil[3]{\small{Department of Physics, University of California at Santa Barbara, CA 93106, U.S.A.}}
\affil[4]{\small{Stanford Institute for Theoretical Physics, 
Stanford University, Stanford, CA 94305, U.S.A.}}
\affil[5]{\small{Department of Physics, Wellesley College, 106 Central Street, Wellesley, MA 02481, U.S.A.}}

\begin{document}

\maketitle

\begin{abstract}
Black holes have an enormous underlying space of microstates, but universal macroscopic physics characterized by mass, charge and angular momentum as well as a causally disconnected interior.   This leads two related puzzles: (1) How does the effective factorization of interior and exterior degrees of freedom emerge in gravity?, and  (2) How does the underlying degeneracy of states wind up having a geometric realization in the horizon area and in properties of the singularity?    We explore these puzzles in the context of an incipient black hole in the AdS/CFT correspondence, the microstates of which are dual to half-BPS states of the $\mathcal{N}=4$ super-Yang-Mills theory. First, we construct a code subspace for this black hole and show how to organize it as a tensor product of a universal macroscopic piece (describing the exterior), and a factor corresponding to the microscopic degrees of freedom (describing the interior).  We then study the classical phase space and symplectic form for low-energy excitations around the black hole. On the AdS side, we find that the symplectic form has a new physical degree of freedom at the stretched horizon of the black hole, reminiscent of soft hair, which is absent in the microstates. We explicitly show how such a soft mode emerges from the microscopic phase space in the dual CFT via a canonical transformation and how it encodes partial information about the microscopic degrees of freedom of the black hole.
\end{abstract}

\pagebreak

\tableofcontents

\pagebreak

\section{Introduction}
Black holes are mysterious objects. In general relativity, one encounters them as solutions to the Einstein equations, but with several peculiar features. These solutions have a spacelike singularity, which is, however, hidden from an external observer by a horizon. What is more, the horizon manifests several thermodynamic properties, where the area of the horizon plays the role of \emph{entropy} as per the Bekenstein-Hawking formula:
\beq
S_{BH} = \frac{A}{4G_N}.
\eeq
Within general relativity, there appears to be no explanation for what this entropy counts.  Modern insights from string theory \cite{Strominger:1996sh}, and in particular the AdS/CFT correspondence \cite{Maldacena:1997re, Witten:1998qj, Gubser:1998bc}, suggest an elegant resolution for this puzzle. In this context, the black hole is realized as a low-energy gravity description of a configuration of D-branes, via the open-closed string duality. A very large number of heavy, almost degenerate, microscopic states of the world-volume theory on the D-branes ($\mathcal{N}=4$ Super Yang Mills, for instance) correspond to essentially the same universal  gravitational geometry, and so the entropy of the black hole simply counts these \emph{microstates}. Furthermore, from the perspective of almost all low-energy probes, these heavy states effectively look like a universal mixed state, namely the thermal ensemble, thus explaining the thermodynamic nature of black holes. 

What are we to make of the singularity in the black hole geometry? This issue was considered in \cite{Balasubramanian:2005mg} in the context of a singular geometry referred to as the half-BPS \emph{superstar} solution \cite{Myers:2001aq}, which is the universal dual geometry corresponding to heavy excited states of fixed charge in the half-BPS sector of $\mathcal{N}=4$ Super-Yang-Mills theory. (Similar questions were also considered previously in the D1-D5 system in \cite{Lunin:2001jy, Lunin:2002qf, Mathur:2005zp}.) We may think of this geometry as an incipient black hole in which the horizon coincides with the singularity -- adding some energy would produce a finite area horizon.
In this example, it can be shown that there is a large class of CFT microstates, which under the AdS/CFT duality correspond to perfectly regular, albeit topologically complex geometries on the gravity side.  Far away from the core, these geometries all effectively look like the superstar, but close to the core they are not singular. Of course, not all CFT microstates need have a smooth geometric interpretation at short distances. Indeed, the vast majority of microstates will have Planck scale features in the dual description.  Thus, close to the core they will not necessarily correspond to smooth geometries, but rather to  some sort of ``spacetime foam''.  
 A classical observer will only have access to a \emph{coarse-grained} description of these states because low-energy probe operators cannot distinguish between different microstates.  Thus, these observers  effectively see a universal mixed state, which for  microstates of a superstar with fixed charge turns out to be the high-temperature thermal ensemble with a fixed number of D-branes.  This ensemble has a dual description as a singular superstar geometry.  In \cite{Balasubramanian:2005mg}, it was argued that this is the origin of black hole singularities in general relativity -- the black hole geometry corresponds to a classical, long-distance description of a large number of underlying microstates which breaks down at the singularity. In the full quantum theory, this singularity gets replaced by a spacetime foam. 

In this way, the AdS/CFT correspondence in principle suggests a resolution to many puzzles about black holes. However, several mysterious features still remain to be understood. In this paper, we will consider two such puzzles in the context of the half-BPS superstar: 

1. How does the effective factorization into microscopic, or ``interior'' degrees of freedom, and  macroscopic, or ``exterior'' degrees of freedom, emerge in gravity? This question is related to the problem of identifying the subspace of the total Hilbert space within which supergravity is a valid description. In  modern AdS/CFT parlance, this subspace is referred to as the \emph{code subspace}, in reference to its connection with quantum error correcting codes \cite{Almheiri:2014lwa}. We will construct a code subspace for the microstates of the half-BPS superstar and argue that it can be approximately organized as a tensor product between coarse-grained ``exterior'' degrees of freedom and fine-grained ``interior'' degrees of freedom.  We show that this factorization is a natural consequence of the \emph{universality} of correlators, namely, the fact that correlation functions of a sufficiently small number of low-energy operators in black hole microstates are universally reproduced by those in a certain thermal ensemble. While this factorization has been anticipated in previous work \cite{Papadodimas:2012aq}, here we will describe a more detailed mechanism for its origin in the context of our incipient black hole. 
    
2. How does the microscopic entropy of the black hole get reflected in gravity in terms of geometric quantities such as the area of the horizon? In other words, does gravity admit degrees of freedom which may encode information about the black hole microstates? Here, we will proceed by studying the classical phase space and its attendant symplectic form for Type IIB supergravity around the superstar geometry.\footnote{The superstar geometry does not have a macroscopic horizon, but for our purposes it is natural to put in a stretched horizon.} For regular half-BPS geometries, this problem was studied in \cite{Grant:2005qc, Maoz:2005nk}, where it was shown to exactly match with the phase space in the dual CFT description. However, a novel feature of the superstar is that it is a singular geometry. This will lead us to consider a regularized gravitational phase space by putting a stretched horizon slightly away from the singular region. While this might seem like a technical detail, it actually has a significant consequence -- we find a new physical mode at this horizon, which would have been a pure gauge degree of freedom in an exact microstate. In detail, if we have access to a precise microstate geometry (assuming that a geometric description exists), then this \emph{soft mode} gets fixed by requiring regularity in the core where the geometry smoothly caps off. The cutoff at the stretched horizon can be understood as a gravitational coarse-graining, which makes these soft modes dynamical. Similar soft modes (or large gauge transformations) have also been encountered in studies of entanglement in gauge theories \cite{Buividovich:2008gq, Ghosh:2015iwa, Donnelly:2015hxa, Donnelly:2016auv, Fliss:2017wop}, where one introduces a fictitious boundary to define a gauge invariant phase space and they make significant contributions to entanglement entropy (see also \cite{Harlow:2016vwg, Lin:2017uzr, Lin:2018xkj} for related discussion in the context of the Ryu-Takayanagi formula).  Relatedly, ``soft hair'' modes have recently also attracted  attention in the context of the black hole information puzzle \cite{Hawking:2016msc, Donnay:2015abr, Donnay:2016ejv, Haco:2018ske}, but their CFT origin in AdS/CFT has remained  unclear. In the present situation we can understand the emergence of such soft modes more comprehensively from the CFT point of view, i.e., we will construct and explicitly coarse-grain the CFT phase space and show the emergence of the soft mode in the infrared as a canonical transformation of the ultraviolet modes. This suggests an explanation for how information about the microstates gets imprinted on the black hole horizon.

The rest of the paper is organized as follows. In  Sec.~\ref{prelim}, we review background material, and in Sec.~\ref{UCS}, we construct the universal code subspace for the 1/2 BPS superstar.  We will argue that this code subspace factorizes into interior and exterior degrees of freedom. In Sec.~\ref{SFGrav}, we  study the gravitational phase space around the superstar geometry and find a new physical mode at the horizon. In Sec.~\ref{SFCFT}, we  construct the phase space from the CFT point of view, and explicitly construct a canonical transformation to extract the effective infrared modes.  These analyses provide a microscopic explanation for the emergence of the macroscopic gravitational soft mode. Technical details are collected in the appendices.

\section{The half-BPS sector of ${\cal N}=4$ SYM}\label{prelim}
\newcommand{\cN}{\mathcal{N}}

\begin{figure}[t]
\centering
\includegraphics[height=4.5cm]{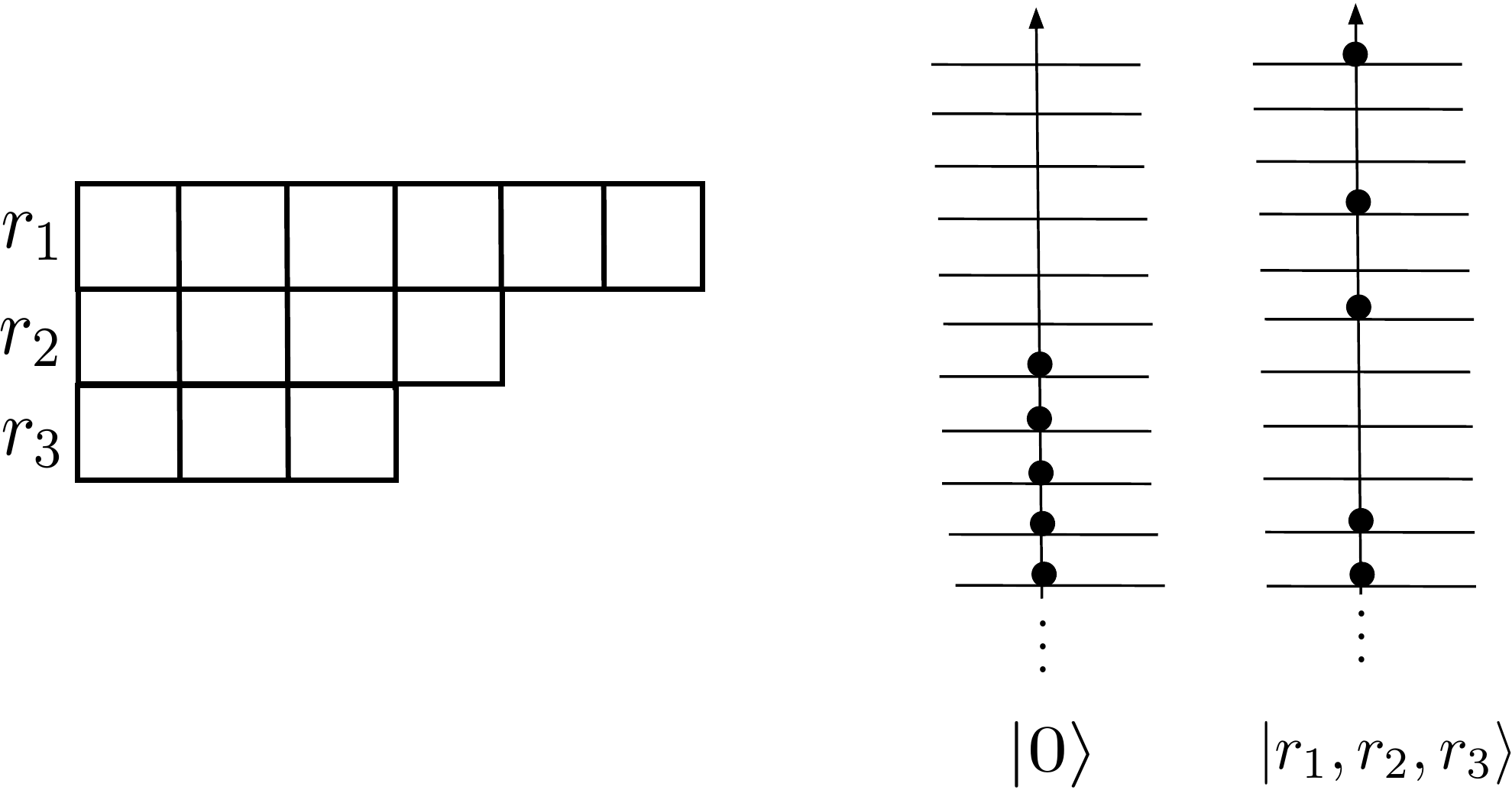}
\caption{\textsf{\small{(Left) A sample Young tableau with row-lengths $(r_1,r_2,r_3)$. (Right) The row-lengths $r_i$ represent excitation energies of the fermions with respect to the vacuum.\label{Tableau} }}}
\end{figure}

The dynamics of the $\frac{1}{2}$-BPS sector of $\cN=4$ Super Yang-Mills (SYM) theory in four dimensions can be reduced to a gauged Hermitian matrix model, as explained in \cite{Berenstein:2004kk, Corley:2001zk} (see also \cite{Berenstein:2017abm}):
\beq \label{MMaction}
L = \frac{N}{2} \int dt\,  \mathrm{Tr}\left((D_{t}X)^2 - X^2\right).
\eeq
Here $X$ is an $N\times N$ Hermitian matrix, which is the S-wave mode of one of the scalars in the $\cN=4$ SYM scalar multiplet, and $D_tX = \pa_tX + i \left[A_0,X \right]$ is the gauge-covariant derivative. We may regard the gauge field $A_0$ as a Lagrange multiplier enforcing the $U(N)$-singlet constraint
\beq
J_{U(N)} = \left[X, \dot{X}\right] = 0,
\eeq
on the Hilbert space, i.e., physical states will be $U(N)$ invariant. This allows us to rewrite the matrix model entirely in terms of the eigenvalues $\{\lambda_i\}$ of the matrix in a harmonic oscillator potential 
\beq
L = \frac{N}{2} \int dt \sum_{i=1}^N \left( \dot \lambda_i^2 - \lambda_i^2\right),
\eeq
with the additional constraint that the $N$-particle wavefunctions are completely antisymmetric under exchange of two eigenvalues. This antisymmetry is required because in the matrix description our states are normalized with respect to the measure 
\beq
\prod_{i,j} dX_{ij} = \prod_i d\lambda_i \Delta(\lambda_i)^2 d\Omega_{U(N)},
\eeq
where $\Delta(\lambda_i) = \prod_{i< j} (\lambda_i - \lambda_j)$ is the Vandermonde determinant and $d\Omega_{U(N)}$ is the Haar measure on $U(N)$. If we want the states $\psi(\lambda_i)$ to be normalized with respect to the flat measure $\prod_i d\lambda_i$, we must absorb the Vandermonde determinant $\Delta(\lambda)$ into the wavefunction, which in turn makes the wavefunctions antisymmetric under exchange of eigenvalues. Therefore, the model reduces to $N$ \emph{fermions} in the simple-harmonic oscillator potential. The energy eigenstates of the model are simply given by antisymmetrized $N$-fold tensor products of the energy eigenstates $|n_i\rangle$ of a single harmonic oscillator. More explicitly, if we label the states by decreasing integers $n_1 > n_2 > n_3 \cdots > n_N \geq 0$, then an eigenstate is given by
\beqn \label{state1}
|n_1,\cdots n_N\rangle_A &=& |n_1\rangle_1 \wedge |n_2\rangle \wedge \cdots \wedge |n_N\rangle\nonumber\\
&\equiv &\frac{1}{\sqrt{N!}} \sum_{P} (-1)^P|n_{p_1}\rangle_1 \otimes |n_{p_2}\rangle_2 \cdots \otimes |n_{p_N}\rangle_N,
\eeqn
where the subscript $A$ stands for anti-symmetrization, and the sum is over all permutations of the $N$ integers. Provided we adhere to our non-decreasing convention, then the states are normalized as
\beq
_A\langle n'_1,\cdots n'_N | n_1,\cdots n_N \rangle_A = \delta_{n_1,n_1'}\cdots \delta_{n_N,n_N'} .
\eeq
The ground state is given by filling the first $N$ levels of the oscillator $|0\rangle = |N-1, N-2, \cdots, 0\rangle_A $, which we may refer to as the Fermi sea. Instead of specifying the list of integers $(n_1,\cdots, n_N)$, we can equivalently specify the list 
\beqn
r_1 &=& n_1-N+1,\nonumber\\
 r_2 &=& n_2-N+2, \\
 &\vdots& \nonumber\\
 r_N &=& n_N.\nonumber
 \eeqn 
Note that the new sequence $r_i$ is non-increasing, i.e. $r_i \geq r_{i+1}$. We can  conveniently represent this information in the form of a \emph{Young tableau}, where the $r_i$s correspond to the row-lengths of the tableau (see Fig.~\ref{Tableau}). In the language used in the matrix model literature, this is the \emph{open string} representation of the Hilbert space. 

Let us next consider the creation and annihilation operators $(\beta_i, \beta_i^{\dagger})$ corresponding to the eigenvalues (i.e., $\beta_i = \frac{1}{\sqrt{2}}(\lambda_i + i \dot{\lambda}_i)$ etc.), which satisfy the usual commutation relations
\beq
\left[\b_i, \b_j^{\dagger} \right] = \delta_{ij},\;\;\;\left[\beta_i, \beta_j \right] = \left[\beta^{\dagger}_i, \beta_j^{\dagger}\right]=0.
\eeq
We can define the gauge invariant operators 
\beq
\cf_k = \frac{1}{N^{k/2}}\sum_{i=1}^N \beta_i^k,\;\;\; \cf_k^{\dagger} = \frac{1}{N^{k/2}}\sum_{i=1}^N {\beta_i^{\dagger}}^k,
\eeq
where 
the sum on eigenvalues imposes the $U(N)$ gauge invariance. We can write these operators in terms of powers of the original matrix, 
\beq \label{CSO}
\cf_k =\frac{1}{N^{k/2}} \mathrm{Tr}\left(\frac{X+i\dot{X}}{\sqrt{2}}\right)^k.
\eeq 

It is a straightforward exercise to check that in the large $N$ limit, when evaluated around the \emph{vacuum}, the operators defined above satisfy the algebra
\beq\label{com}
\left[ \cf_k, \cf_m^{\dagger} \right] = k\delta_{km} + O(k^2/N).
\eeq
The corrections in equation \eqref{com} become important when $k \sim N^{1/2}$. Therefore, this algebra comes with the cutoff $k \leq N^{\frac{1}{2}-\epsilon}$. Further, this algebra is valid in a subspace of the full Hilbert space which is constructed by acting on the vacuum with a sufficiently small number of $t_k^\dagger$ operators. This subspace, which  behaves like the Fock space for the $\cf_k$ operators, can be regarded as a \emph{code subspace} for the algebra around the vacuum state. In \cite{Berenstein:2017rrx}, this code subspace was discussed around special excited states corresponding to rectangular Young tableaus.  In matrix model language, this is a \emph{closed string representation} of the Hilbert space. Note that while the three-point functions of the $\cf_k$s are suppressed by $1/N$ inside the code subspace, they are still non-trivial and represent splitting and joining interactions for closed strings. The transformation between the open string and closed string representations can be elegantly understood in terms of the representation theory of symmetric groups \cite{Berenstein:2017abm}, but we will not need these details here.

\subsubsection*{Phase-space density}
\begin{figure}[!h]
 \centering
 \includegraphics[height=5cm]{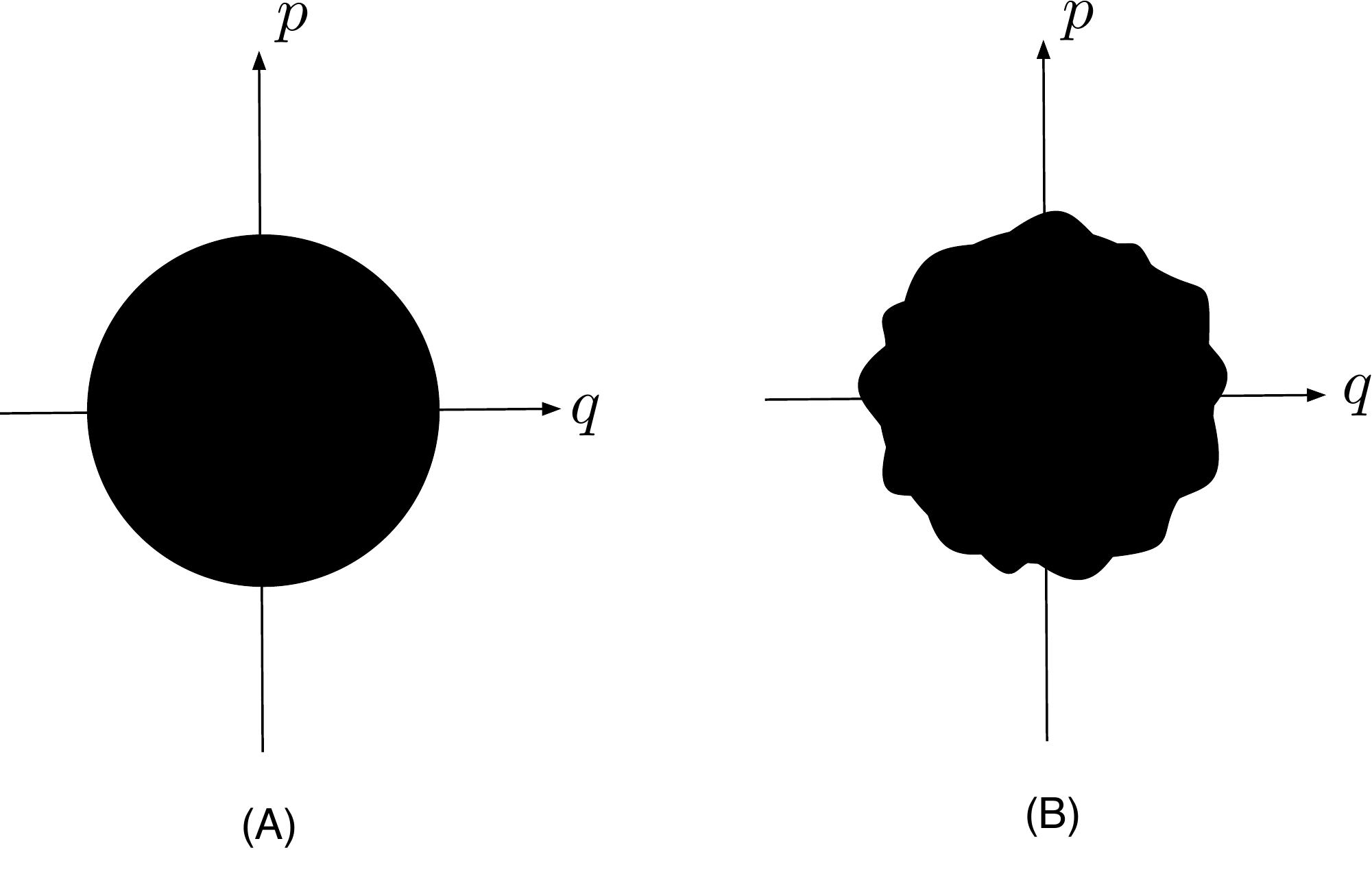}
 \caption{\small{\textsf{(A) The density function for the Fermi sea corresponds to a black disc in the one-particle phase space, whose radius fixes the AdS radius in the gravity dual. (B) Coherent states of the $\cf_k$ operators create shape deformations on the surface of the Fermi sea.\label{vs}}}}
 \end{figure}

It is useful to define the one-particle \emph{phase space density}, also known as the \emph{Wigner density}: 
\beq \label{FD1}
u(q,p) = 2 \int_{-\infty}^{\infty} dr\, e^{\frac{2i p r}{\hbar}} \left\langle q-r | \hat{\rho}_1 | q + r\right\rangle, 
 \eeq
where $\hat{\rho}_1$ is the reduced density matrix obtained by tracing out $N-1$ fermions \cite{Dhar:1992rs,DHAR1996234, Dhar:1993jc, Balasubramanian:2005mg, Balasubramanian:2007zt}, while $q$ and $p$ are the usual phase space coordinates.  We can also express this density function in terms of \emph{second-quantized} fermion creation operators. Let $(\hat\psi_n, \hat\psi_n^{\dagger})$ be anti-commuting operators, which annihilate and create fermions in the (one-particle) energy level $n$. These satisfy the commutation relations
\beq \label{SQFO}
\left\{\hat\psi_m , \hat\psi_n^{\dagger} \right\} = \delta_{mn}, 
\eeq
and we may write the state \eqref{state1} in terms of these as
\beq
|n_1,\cdots, n_N\rangle_A = \hat\psi_{n_1}^{\dagger}\hat\psi_{n_2}^{\dagger} \cdots \hat\psi_{n_N}^{\dagger} |\mathbf{0}\rangle
\eeq
where $|\mathbf{0}\rangle$ denotes the second-quantized vacuum. In terms of the $\psi_m$s, the density operator is given by
\beq \label{FD2}
\hat{u}(q,p) = 2  \int_{-\infty}^{\infty} dr\, e^{\frac{2i p r}{\hbar}} \hat{\psi}^{\dagger}( q-r ) \hat{\psi}( q + r), 
\eeq
where 
\beq
\hat{\psi}(q) = \sum_n f_n(q) \hat\psi_n, 
\eeq
and $f_n(q)$ are the one-particle Harmonic oscillator wavefunctions in position space. The function $u$ defined in equation \eqref{FD1} is simply the one-point function of the operator $\hat{u}$ in the appropriate $N$-particle state. The function $u$ in the classical limit, $N \to \infty,\;\hbar \to 0$ with $N\hbar$ fixed, can be interpreted as a density for fermion occupation in the one-particle phase space satisfying 
\beq
\int \frac{dpdq}{2\pi \hbar}\, u(p,q) = N,
\eeq
and plays a crucial role in the correspondence with the gravitational dual geometry (see Sec.~\ref{SFGrav}).\footnote{On the gravity side of the AdS/CFT correspondence, the classical limit corresponds to fixing the AdS radius $\ell_{AdS}^4 = 2N\hbar$, while sending the Planck length $\ell_P \to 0$.} For example, in the classical limit the phase-space density corresponding to the Fermi sea is given by $u(q,p) = \Theta(q^2+p^2 - 2 \hbar N)$. We can pictorially represent this as a black disc in the one-particle phase space (see Fig.~\ref{vs}), where the black region corresponds to $u=1$ and the white region corresponds to $u=0$. The action of the $\cf_k$ operators can be thought of as creating periodic waves on the surface of the Fermi sea, while a localized shape perturbation of the Fermi sea can be interpreted as a coherent state in terms of $\cf_k$s. 

For later use, we also record the commutation relations satisfied by $\hat{u}$ \cite{Dhar:1992rs}. These can be obtained by using the defining equation \eqref{FD2}:
\beqn
\left[ \hat{u}(p,q), \hat{u}(p',q')\right]&=&4\int drdr'e^{\frac{2i(rp+r'p')}{\hbar}}\left\{\hat{\psi}^{\dagger}(q - r)\hat{\psi}(q'+r') \delta(q-q'+r+r')\right\}- (q\leftrightarrow q') \nonumber\\
&=&4\int drdr'e^{\frac{2i(rp+r'p')}{\hbar}}e^{r'\pa_q - r\pa_{q'}}\left\{\hat{\psi}^{\dagger}(q - r_+)\hat{\psi}(q'+ r_+) \delta(q-q')\right\}- (q\leftrightarrow q') \nonumber\\
&=&2i \sin\left\{\frac{\hbar}{2}(\pa_{p}\pa_{q'} - \pa_{p'}\pa_{q})\right\}\Big(2\pi\delta(p-p') \delta(q-q')\hat{u}(p,q)\Big),\label{CR0}
\eeqn
where in the second line we have used $r_+ = r+r'$. If we define the operator $\hat{T}_{\pi} =\int \frac{dp dq}{2\pi\hbar} \pi(p,q) \hat{u}(p,q)$ where $\pi$ is an arbitrary function on phase space, then in the $\hbar \to 0$ limit the above commutation relations imply
\beq
\left[\hat{T}_{\pi_1}, \hat{T}_{\pi_2}\right] = \hat{T}_{\{\pi_1, \pi_2\}_{PB}},
\eeq
where $\{\pi_1, \pi_2\}_{PB} = \epsilon^{ij}\pa_i\pi_1\pa_j\pi_2$ are the classical Poisson brackets in the one-particle phase space. Thus, in the $\hbar \to 0$ limit, $\hat{u}$ generates the Poisson algebra on phase space. As we will see later, we are interested in these smeared operators because they generate the subspace of low energy states, which is in one to one correspondence with long wavelength excitations of the bulk.

\section{Universal code subspaces and incipient black holes}\label{UCS}

The half-BPS sector of $\mathcal{N}=4$ super-Yang Mills includes very heavy states created by operators with conformal dimension $\Delta \sim O(N^2)$ that are dual to extremal black holes called ``superstars'' \cite{Myers:2001aq, Balasubramanian:2005mg, Suryanarayana:2004ig} (see \cite{Simon:2018laf} for a recent discussion).  Below we will describe this ``superstar ensemble'' of states, and later explain in Sec.~4  their coarse-grained description in the gravity dual in terms of a singular metric.   We will then analyze the structure  of states in the Hilbert space which are ``close'' to typical states in the superstar ensemble. This will constitute a ``code subspace'' \cite{Almheiri:2014lwa, Papadodimas:2012aq} of  states created by the action of universal, low-energy operators on typical states of the ensemble. The universality of the code subspace leads to an  effective factorization into  macroscopic ``exterior'' and microscopic ``interior'' degrees of freedom. 

\subsection{Superstars}
\begin{figure}[!h]
 \centering
 \includegraphics[height=4cm]{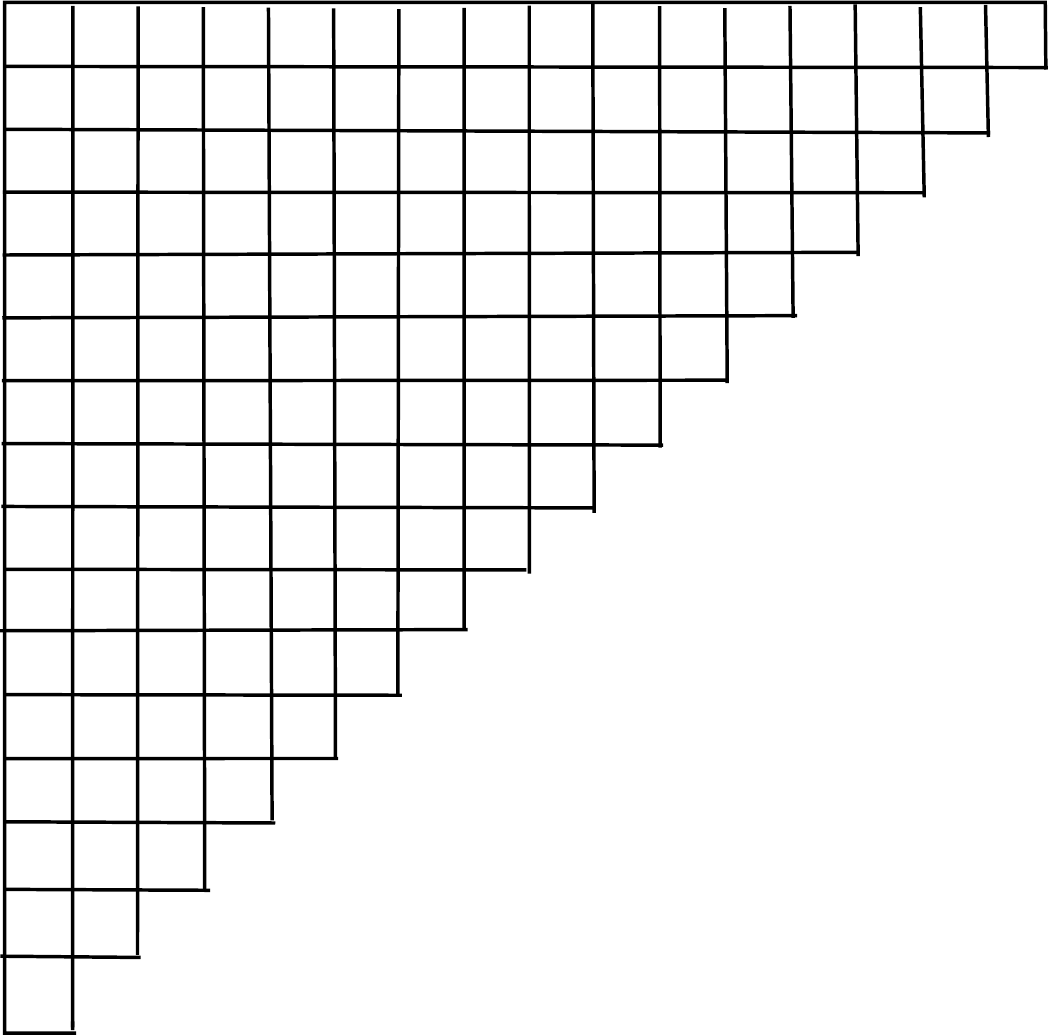}
 \caption{\small{\textsf{A microstate in the superstar ensemble is a triangular tableau, here shown for $\omega =1$. \label{TrTableau}}}}
 \end{figure}

We will be interested in states  of energy $\Delta = O(N^2)$ above the Fermi sea. As we will describe in Sec.~\ref{SFGrav}, the universal, coarse-grained gravitational description of these states is an incipient black hole called the \emph{superstar}. From the CFT point of view, a typical state of this type lives in an energy window around a \emph{triangular tableau} with $N$ rows and $N_c = \omega N$ columns (Fig.~\ref{TrTableau}), i.e., it differs from such a triangular tableau by a small number of boxes \cite{Balasubramanian:2005mg}. Here ``small'' refers to having the same average slope $\omega$ of the Young tableau and approximately equal energy; since the triangular tableau with $N$ rows and $\omega N$ columns has order $N^2$ boxes, it is possible for a second state to differ in $O(N)$ boxes and still be part of the same microcanonical ensemble.
 
 It is convenient to describe these states by introducing the following ensemble: 
\beq \label{supens}
\rho_{\star} = \frac{1}{Z_{\star}}\sum_{c_1,\cdots,c_N =0}^{\infty}e^{-\beta\sum_{j=1}^Njc_j - \mu\left(\sum_jc_j - N_c\right)} |c_1,\cdots, c_N\rangle \langle c_1,\cdots, c_N|
\eeq
where following \cite{Balasubramanian:2005mg}, we have introduced the new variables $c_N = r_N, \; c_{i} = r_{i}- r_{i+1}$. The variable $c_i$ counts the total number of columns of length $i$ in the tableau. 
Here $\beta$ is the inverse ``temperature'' (which multiplies the energy), while $\mu$ is a Lagrange multiplier which enforces the constraint that the total number of columns is $N_c$.\footnote{On the gravity side, this constraint is dual to fixing the number of giant gravitons wrapping an $S^3$ inside the $S^5$ \cite{McGreevy:2000cw, Balasubramanian:2001nh}.}
The parameters $q = e^{-\beta}, \;\zeta = e^{-\mu}$ are determined by the equations 
\beq
\sum_{j=1}^N \frac{j\zeta q^j}{1- \zeta q^j} = \Delta,\;\;\sum_{j=1}^N \frac{\zeta q^j}{1- \zeta q^j} = N_c,
\eeq
with $\Delta$ being the energy. In the $\beta \to 0$ limit, the above equations simplify greatly, and we obtain
\beq
\Delta = \frac{N_c(N+1)}{2},\;\; \frac{\zeta}{1-\zeta} = \frac{N_c}{N} \equiv \omega.
\eeq
It was argued in \cite{Balasubramanian:2005mg} that this limit leads to a universal description of almost all superstar microstates.  In this limit, the average value of $c_k$ is given by $\langle c_k \rangle_{\star} = \omega$. This implies that the typical state in the ensemble lies very close to a triangular tableau, where $\omega = N_c/N$ determines the slope of the tableau (Fig.~\ref{TrTableau}).  Correlators of low-energy operators in a typical triangular tableau state universally reproduce correlation functions of such operators in almost all states of energy $\Delta$.  This is essentially ``eigenstate thermalization'' of BPS states in the large $N$ limit. We will henceforth refer to equation \eqref{supens} in the $\beta \to 0$ limit as the superstar ensemble. The entropy of this ensemble scales linearly with $N$ and is given by 
\beq \label{SSentropy}
S = -\ln\left(\frac{\omega^{N_c}}{(1+\omega)^{N+N_c}}\right).
\eeq

\begin{figure}[!h]
 \centering
 \includegraphics[height=3cm]{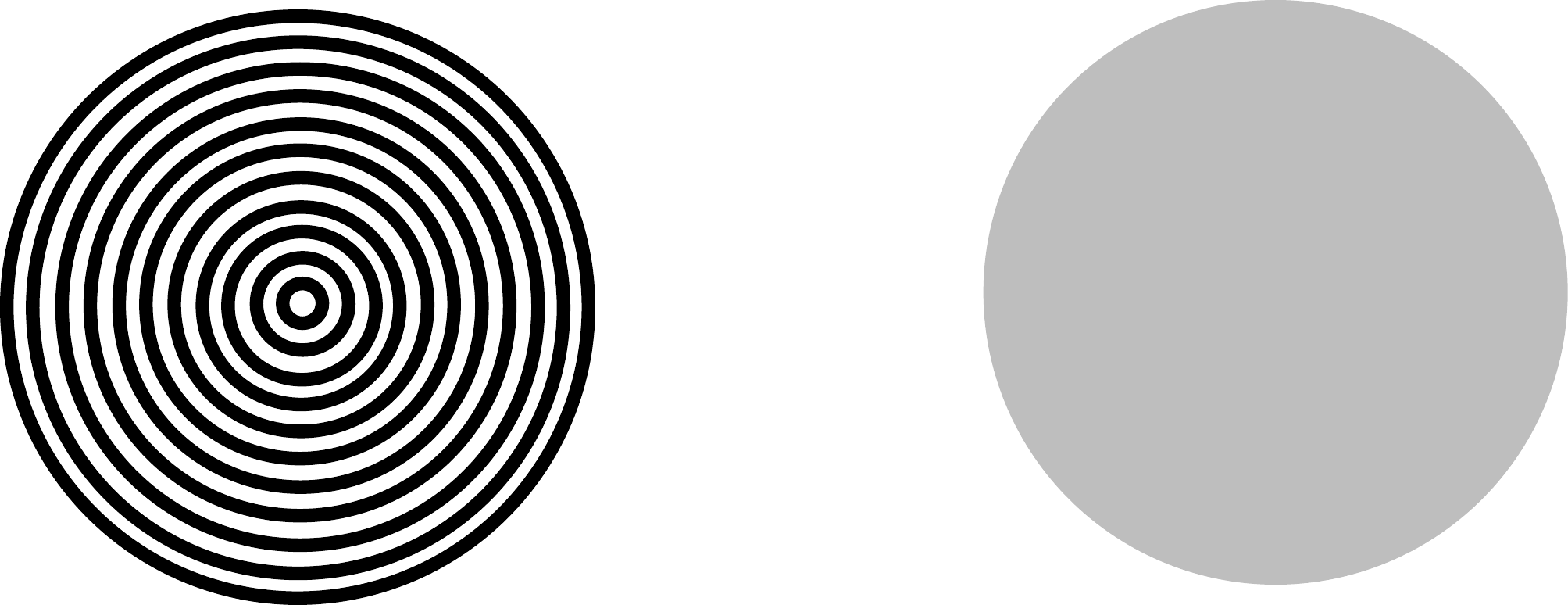}
 \caption{\small{\textsf{(Left) We naively expect a typical microstate in the superstar ensemble (here with $\omega$ =1) to be composed of a dense set of concentric black and white rings. (Right) The coarse grained density looks like a grey disc. \label{figCG}}}}
 \end{figure}

In terms of the density function $u(q,p)$ introduced previously, we naively expect the typical superstar state to look like a large number of black and white rings, each with area $\hbar$.  This is because each excited fermion will move periodically in the oscillator potential with a frequency determined by its energy -- in phase space, this corresponds to a density function that occupies a ring of area $\hbar$ and squared radius equal to the energy.  Further, the rings corresponding to different fermions will be separated by gaps of area $O(\hbar)$.  More precisely, the quantum phase space density oscillates between concentric black and white rings representing occupied and unoccupied regions, with an average amplitude determined by $\omega$ and oscillation frequency at the $\hbar$ scale in phase space:\footnote{Here and below, by oscillations at the $\hbar$ scale we are referring to the area-scale in the one-particle phase space over which the density fluctuates.}
\beq
u(q,p) = \frac{1}{1+\omega}\Theta\left(p^2+q^2-2N\hbar(1+\omega)\right)+ {\rm oscillations \ at \ the} \ \hbar \ {\rm scale}. \,
\eeq
Such a density does not have a natural classical limit ($N\to \infty$, $\hbar \to 0$ with $N\hbar$ fixed), because of the above fluctuations, which do not disappear in this limit. However, a classical observer will only have access to a \emph{coarse-grained} description of this density, which corresponds to averaging over it at some scale $y_0 \gg \hbar^{1/2}$. Therefore, the coarse-grained density will essentially be a grey disc whose radius is fixed by $\Delta$ and whose greyness is determined by the slope of the triangular tableau $\omega = N_c/N$ (Fig.~\ref{figCG}, \cite{Balasubramanian:2005mg, Balasubramanian:2007zt}):
\begin{equation}
    u(q,p)=\frac{1}{1+\omega}\Theta\left(p^2+q^2-2N\hbar(1+\omega)\right).
\end{equation}
We will develop this in more detail in
 Sec.~\ref{SFGrav}; we will explain there how it leads to the emergence of a singularity in the universal gravitational geometry dual to the superstar microstates, and further results in novel and interesting features in the phase space of excitations around this geometry.  

\subsection{Code subspaces around black holes -- a general picture}\label{sec:genpic}

We want to analyze the structure of the Hilbert space of superstar microstates in the vicinity of the typical configurations in the ensemble. In this section, we will first present arguments applicable in a more general context, and then specialize to the superstar in the subsequent sections. To this end, consider a set of states of energy within a microcanonical energy window around $\Delta \sim  O({N^2})$.  These states will span an $\sim e^{S}$ dimensional subspace ${\cal H}_{\Delta}$ of black hole microstates, where $S$ is the entropy of the black hole.
We are going to consider a set of $e^{\varepsilon S}$ {\it reference states} with $\varepsilon<1$, which we will denote by ${\cal H}_{\varepsilon,\Delta}$.
${\cal H}_{\varepsilon,\Delta}$ has a much smaller dimension than ${\cal H}_{\Delta}$, and thus, if we pick two random states in this subspace $|\alpha\rangle,|\alpha'\rangle$, they will be orthogonal.\footnote{More precisely, if we pick two random states as superpositions of Young tableux, then their overlap will be $e^{-N}$ because of dephasing.  We ignore these small corrections.}   Next, on top of each reference state, we will consider  small excitations by the action of low-energy operators, where ``small'' means that the perturbations of the different reference states remain orthogonal.  
(In the 1/2 BPS case that we are studying, examples of such operators will be $(t_k, t_k^{\dagger})$ with $k \ll \sqrt{N}$.) Thus $\mathcal{H}_{\varepsilon, \Delta}$ has the structure:
\begin{equation}
    {\cal H}_{\varepsilon,\Delta}=\oplus_{\alpha} {\cal H}_{\alpha,small}, \label{alpha}
\end{equation}
where each of the ${\cal H}_{\alpha,small}$ represents the small subspace formed by acting on the reference state $|\alpha\rangle$ with small excitations and has a dimension that is less than $O(\sqrt{N})$ (i.e.,  states in this Hilbert space differ from the reference Young tableau $\alpha$ by less than $\sqrt{N}$ boxes).

Given a reference state for each $\alpha$ in equation \eqref{alpha}, we expect that a natural notion of universal, low energy excitation should involve operators ${\cal O}_{simple}$ which cannot  precisely distinguish the microstate in question (see \cite{Balasubramanian:2005mg,Balasubramanian:2006jt} for a discussion in the present LLM context). In other words, 
\begin{equation}
    \langle \alpha'|{\cal O}_{simple}|\alpha\rangle=\delta_{\alpha,\alpha'} \langle {\cal O}_{simple}\rangle_{\Delta} +\cdots \label{ETHlike}
\end{equation}
Here $\langle {\cal O}_{simple}\rangle_{\Delta}$ only depends on the energy $\Delta$.  The ellipses denote corrections that can be off-diagonal in $\alpha$ and $\alpha'$, and can also depend on quantum numbers other than the energy.  This form also implies that $\mathcal{O}_{simple}$ cannot annihilate the reference state because, if it did, the leading expectation value could not  depend universally on the energy.  This is natural to expect because the black hole microstate is heavy and complex, and thus a light and simple operator cannot possibly annihilate it.

The formula \eqref{ETHlike} is reminiscent of the Eigenstate Thermalization Hypothesis (ETH) \cite{PhysRevA.43.2046, PhysRevE.50.888} (see also \cite{Lashkari:2016vgj} for a recent discussion in the context of CFTs): we expect simple correlators in complex energy eigenstates to behave like  correlators in the thermal ensemble. From the ETH we expect that the off-diagonal corrections  to \eqref{ETHlike} will be  exponentially suppressed in the entropy, while corrections to the diagonal terms will be polynomial in the entropy.   In our case of interest, namely the superstar, the entropy scales as $S \sim N$ (see above), and so the diagonal corrections will be power-law suppressed in $N$; in this way, large $N$ plays the role of the volume in the usual ETH.  Meanwhile, the off-diagonal terms vanish for us when $\varepsilon < 1$ because the reference states have then been chosen to be sufficiently far apart so that there is no way for a simple operator to move the system from one $\alpha$ to another. Indeed this is why the subspace we consider has a direct sum structure.   As we let $\varepsilon$ tend to 1 the direct sum structure will break down
and there will be $O(1)$ off-diagonal terms in
\eqref{ETHlike}.
Based on~\cite{Berenstein:2017abm}, the action of traces is local on the location of the edge of a given Young tableaux and changes very few contiguous boxes in a superposition of all possible locations: if two reference tableaux are very different, it takes a lot of moves by such actions to go from one to the other one.
This property of acting locally on the edge of a diagram ensures the property of equation \ref{ETHlike}.

Due to the diagonal structure of \nref{alpha}, the simple operators act within blocks.  So we can write
\begin{align}
{\cal O}_{simple}=\oplus_{\alpha} {\cal O}_{simple,\alpha}\,.
\end{align}
What is more the ETH-like formula \eqref{ETHlike} tells us that the action of these operators will be universal in that to leading order correlators will only depend on the energy of the reference state \footnote{For other general half BPS states in different ensembles, it should only depend on the gray coloring of the LLM plane at the coarse grained scale.}.   Because of this, we can represent the algebra of simple operators on a universal code subspace $\mathcal{H}_{univ}$  built around any fiducial reference state.   The low-energy operators are then universally represented as some $\mathcal{O}_{univ}$ whose correlators in $\mathcal{H}_{univ}$ reproduce those in ${\cal H}_{\alpha,small}$ for any $\alpha$ up to corrections polynomial in the entropy.   In this sense, to leading order in powers of the entropy, we find that $\mathcal {\cal H}_{\alpha,small} \sim {\cal{H}}_{univ}$ are isomorphic.  Here we have focused on simple, low-energy operators that respect the block structure of $\mathcal{H}_{\varepsilon, \Delta}$.  Of course, maintaining this structure also requires us to truncate the algebra of the operators so that the excitations they create are not too energetic or too complex.

The direct sum structure of ${\cal H}_{\varepsilon,\Delta}$ means that we can write any state as $|\alpha,i_\alpha \rangle$, where $\alpha$ labels what reference state we are working around and $i_\alpha$ labels a state in ${\cal H}_{\alpha,small}$. But correlation functions can be computed up to subleading corrections by working with $|\alpha \rangle \otimes |i\rangle$ where $\alpha$ is now a super-selection sector label telling us what reference state we are working around, $|i\rangle \in \mathcal{H}_{univ}$, and the action of simple operators is $\mathcal{O}_{simple} = \mathbb{I}_{SS} \otimes \mathcal{O}_{univ}$. In other words, 
\begin{equation}
    {\cal H}_{\varepsilon,\Delta} \sim {\cal H}_{SS} \otimes  {\cal H}_{univ} \,,
\end{equation}
where ${\cal H}_{SS}$ is the super-selection sector label keeping track of which reference state we are working around.

Above, we have presented a scenario in which a tensor product structure can effectively appear in a complex Hilbert when it is probed exclusively with simple operators.  Below, we will explicitly demonstrate how this picture is realized for the half-BPS superstar of AdS$_5$ gravity.

\subsection{Code subspace around the superstar}\label{LEC}


In the notation of the Sec.~\ref{sec:genpic}, $|\alpha\rangle$ will be a typical reference microstate of the superstar.  As we discussed, this will be a Young tableau state of excitation energy $\Delta$ lying close to the triangular tableau.  Since we are picking the reference states at random, they will generically differ by more than $O(\sqrt{N})$ boxes, which will allow us to consider the orthogonal code subspaces that we described above.   
Any such reference state will have a phase space density $u$ \cite{Balasubramanian:2005mg, Balasubramanian:2007zt}
\beq
\langle \alpha | \hat{u}(q,p) | \alpha \rangle= u_0\Theta(p^2+q^2-R^2)+ {\rm oscillations \ at \ the} \ \hbar \ {\rm scale} \,
\eeq
where $u_0 = \frac{1}{1+\omega}$, $R^2 = (1+\omega) 2N\hbar$, and recall that the operator $\hat{u}$ was defined in equation \eqref{FD2}.   Coarse-graining at a scale bigger than $\hbar^{1/2}$ averages over the oscillations and should lead to $\hbar$ suppression of differences in the phase space density between reference microstates. In the classical limit, $\hbar \sim 1/N$, so these suppressions are the power law in the entropy suppressions we have discussed in section $3$.

Small perturbations of the phase space density provide a natural candidate class of low-energy fluctuations.   To this end,  consider the  coherent states:
\beq\label{coadorb}
|\psi\rangle = e^{i\hat{T}_{\pi}} |\alpha\rangle,\;\;\;\hat{T}_{\pi} = \int \frac{dp'dq'}{2\pi\hbar} \pi(q',p')\hat{u}(q',p'),
\eeq
where $|\alpha\rangle$ is a reference microstate and $\pi$ is a function on phase space.  Infinitesimally this reads
\beq \label{preinfCO}
\delta|\psi\rangle = i\int \frac{dp'dq'}{2\pi\hbar} \delta \pi(q',p')\hat{u}(q',p') |\alpha\rangle. 
\eeq
 Using the commutation relations for $\hat{u}$ \nref{CR0} and taking the $\hbar \to 0$ limit, we find that the infinitesimal change in the one-point function implied by the change in the state \eqref{preinfCO} is 
\beq\label{predu-app}
\delta_{\pi} 
\langle \hat{u} \rangle = \epsilon^{ij}\pa_i \delta \pi \pa_j u = \left\{\delta \pi, u \right\}_{PB}. 
\eeq
Although we defined the state deformation in terms of the function $\delta \pi$, we can invert  \eqref{predu-app} to solve for $\delta \pi$ in terms of the change in the expectation value $\delta u$.  So we can equivalently label state deformations by $\delta u$.  
If $\delta u(p,q)$ is sufficiently coarse (with respect to the $\hbar^{1/2}$ scale), the deformation will behave universally on all microstates because, as we discussed above, the difference in the coarse-grained reference state densities will be $1/N$ suppressed.

It is helpful to expand the $\hat{T}_\pi$ operators as
\begin{equation}
    \hat{T}_\pi = \sum_{k,l} \pi_{k,l} t_{k,l} ~~~{\rm where}~~~
     \pi(p,q) =\sum_{k,l} \pi_{k,l} \left(\frac{q-ip}{\sqrt{2N\hbar}}\right)^k \left(\frac{q+ip}{\sqrt{2N\hbar}}\right)^l
\end{equation}
Then by definition
\begin{equation}
    t_{k,l} =\int \frac{dp dq}{2 \pi \hbar  (2N \hbar)^{(k+l)/2}} (q-i p)^k (q+i p)^l \hat{u}(p,q)
    \end{equation}
It can be shown that in terms of the matrices $X$ in the definition of half-BPS sector \cite{DHAR1996234} \eqref{MMaction}
\begin{equation}
    t_{k,l} = \frac{1}{N^{(k+l)/2}} \text{Tr} \bar{Z}^k Z^l ~~~;~~~ Z = {X + i\dot{X} \over \sqrt{2\hbar}} \, .
\end{equation}
The normalization is chosen so that the one point function in the vacuum (with $u=1$ inside a disc of radius $2N\hbar$ and zero outside) is 
$    \langle 0 | t_{k,l} | 0 \rangle =   \frac{ N}{(k+1)} \delta_{k,l}  +O(1) 
 $. For computations it is useful to write these operators in terms of the second quantised fermion operators $\hat\psi_n$  \eqref{SQFO}:
\begin{align}
t_{k,l} = \frac{1}{N^{(k+l)/2}}\sum_{n=0}^\infty \frac{\sqrt{(n+k)!(n+l)!}}{n!} \hat\psi^\dagger_{n+k} \hat\psi_{n+l} \,.
\end{align}
In terms of these modes, the commutation relations \nref{CR0} read (see also \cite{DHAR1996234}):
    \begin{equation}
        [t_{k,l},t_{k',l'}]=\frac{1}{N} (k' l-k l') t_{k+k'-1, l+l'-1} \, .
        \label{tcomms}
    \end{equation}
It is then clear that the $1/N$ corrections to this expression will be important if any of the $k l' \sim O(N)$. This imposes an effective cutoff on the sort of excitations that we should consider as part of the code subspace in order to guarantee that (\ref{tcomms}) act like standard creation and annihilation operators building a space of excitations. In other words, we require that $k,l < O(\sqrt{N})$.  This condition is a precise version of the requirement we made above that the perturbations $\delta \langle \hat{u} \rangle$ should behave universally around all microstates.   Additionally, this guarantees that the action of $O(1)$ $t_{k,l}$ operators will not take the system from one reference microstate to another since these differ by $O(\sqrt{N}$) boxes or more.


\subsection*{Universality}

The code subspace we have defined will be universal if low-order correlation functions of the $t_{k,l}$ are identical up to $1/N$ corrections in reference microstates of the superstar ensemble.  If this is the case, as we discussed above, the Hilbert takes  an approximately factorized form as ${\cal H}_{ref} \otimes {\cal H}_{code}$ when probed by these operators.   We will test  universality  by computing correlators of operators constructed as sums and 
$O(1)$ products of the $t_{k,l}$. We will compare these correlators as computed in the superstar ensemble and in microstates. (Also see \cite{Balasubramanian:2005mg, Balasubramanian:2007qv}.) 

Firstly, we can evaluate the commutator of $t_{k,l}$ operators in the superstar ensemble using  \eqref{tcomms} to obtain:
\begin{equation}
  \langle [t_{k,l},t_{k',l'}]  \rangle_\star =\delta_{k+k',l+l'} \frac{k' l-k l'}{k+k'} u_0^{-(k+k'-1)},
\end{equation}
where recall that $u_0 = \frac{1}{1+\omega}$. Equivalently, we could have evaluated this commutator in a microstate and obtained the same result because of the universality of the one-point function of the $t_{k,l}$s (see the right hand side of equation \eqref{tcomms}). In this way, we get the same commutators for any microstate of the superstar, as long as $k,l$ are not too big.  In other words, the commutator is universal.

\begin{figure}[!h]
\centering
  \begin{tabular}{c c}
  \includegraphics[height=5cm]{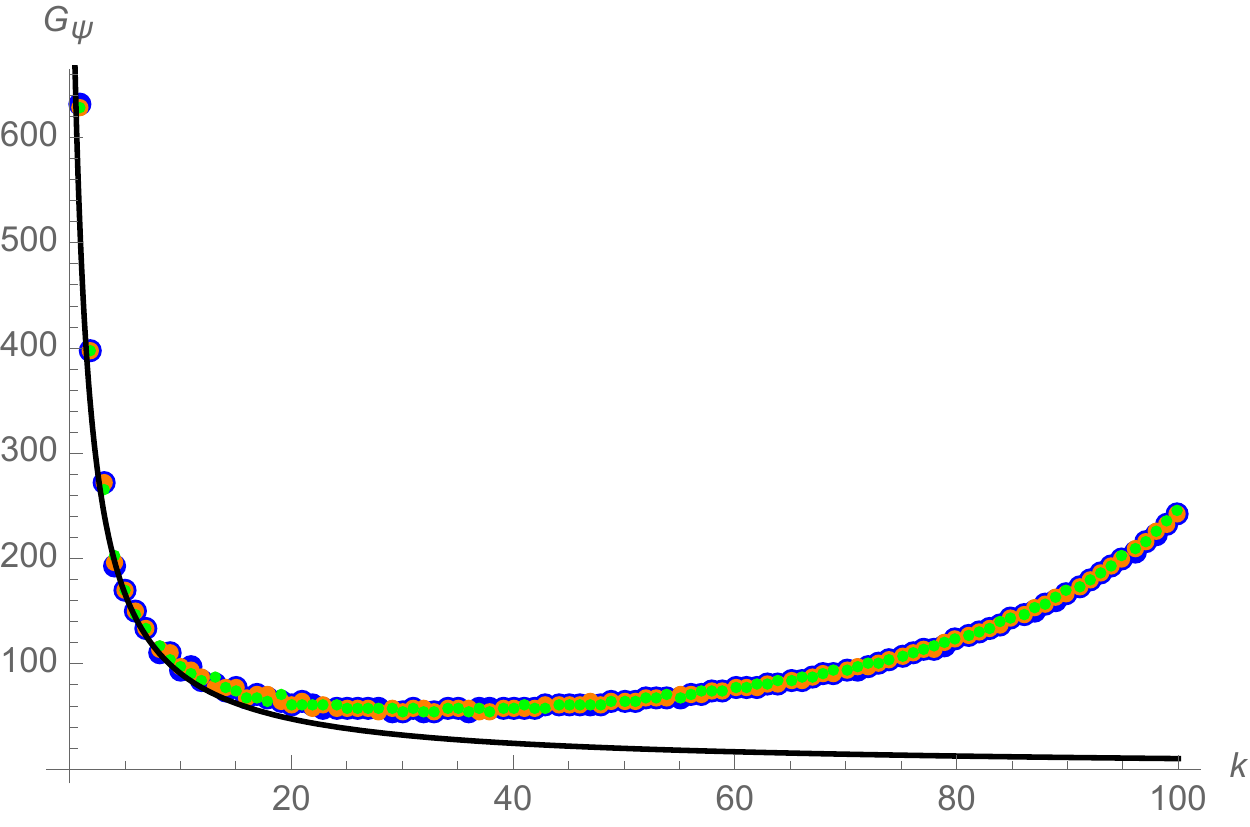} & \includegraphics[height=5cm]{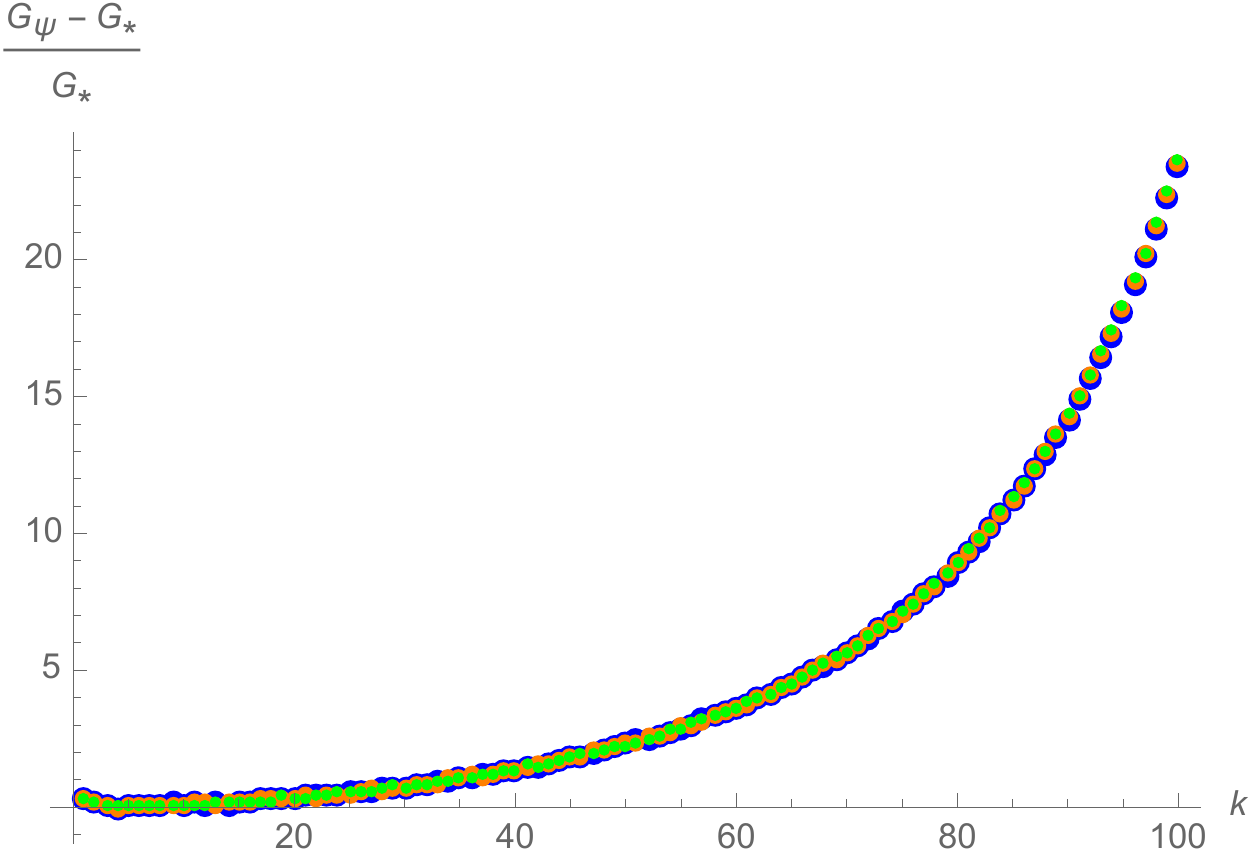}
  \end{tabular}
 \caption{\small{\textsf{(Left) The two-point function $G_{\psi}$ as a function of $k$, for three particular microstates denoted by blue, orange and green points (the points are almost overlapping so we have chosen different point-sizes to make them more visible), which are linear combinations of five tableaus close to the typical one, but chosen randomly. For comparison, we have also shown the large-$N$ result from the superstar ensemble with $\omega =1 $ (black line). (Right) The relative deviation of the two-point function from that in the superstar ensemble as a function of $k$. Here $N = 1000$. \label{ETH} }}}
 \end{figure}

Let us now consider the two-point functions. We can compute the connected two-point functions
analytically in the superstar ensemble \eqref{supens}: 
\begin{align}\label{SS2PF}
  \langle t_{k,l} t_{k',l'}\rangle_{\star} =  N \frac{ (1- u_0) u_0^{-(k+k')}}{1+k+k'} \delta_{k+k',l+l'} +O(N^0)  \, .
\end{align}
We want to compare this formula with the  two point function in a generic microstate
\beq
| \psi \rangle = \sum_{\alpha}\sqrt{ p_{\alpha}} |\{ c^{\alpha}_i \} \rangle,\;\;\;  \sum_{\alpha} p_{\alpha}=1,
\eeq
where $\alpha$ labels states in the superposition. 
If the two-point function of $t_{k,l}$ operators is universal, we expect the microstate calculation to be equal to the result in the superstar ensemble, equation \eqref{SS2PF}.
 Since the computation differs in detail for each microstate, we tested this numerically (Fig.~\ref{ETH}), focusing on correlation functions of $t_k \equiv t_{0,k}$ and $t_k^\dagger = t_{k,0}$ for simplicity. The microstates in question were taken to be uniform superpositions of five randomly chosen tableaus close to the typical one. We see that for $k \ll \sqrt{N}$ the microstate two-point function matches the large-$N$ superstar result closely, while for $k \sim \sqrt{N}$ we see that a correction linear in $k$ kicks in, and then subsequently higher-order corrections appear. This demonstrates that the $\cf_k$s have universal correlators \cite{Hayden} around the superstar, i.e., correlation functions of a sufficiently small number of low-energy operators in a generic eigenstate are universally reproduced by the superstar ensemble.  

 Finally, although the original ${\cal N} = 4$ SYM theory has zero physical temperature, the superstar ensemble effectively acts like a finite temperature system, as we would expect for black hole microstates.
 Finite temperature behavior can be diagnosed from the analytic structure of position space correlators.  Because we are working with a matrix model, the only ``position'' is in time. Fourier transforming the momentum space operators $t_k$ gives
  \beq
  \cp(w) = \sum_{k=1}^{\Lambda} \left(\cf_k w^k + w^{-k} \cf_k^{\dagger} \right),
  \eeq
where $\Lambda \sim N^{1/2-\epsilon}$ is the cutoff on the code subspace that we discussed previously, and $w=w_0 e^{-it}$ (where $w_0$ is a normalization constant). We can compute the two point function of  $\cp(w)$ in the superstar ensemble with $ u_0 = \frac12$ using \eqref{SS2PF}. The leading order result in the $N \to \infty$ limit is 
\beq
  G_{\star}(w) = \langle \cp(w) \cp(0)\rangle_{\star} \propto N\left[-2 -w\ln(1-\frac{1}{w}) - \frac{1}{w}\ln(1- w)\right].
  \eeq
In the complex $w$ plane, this function has branch cuts from $(0,1)$ and $(1, \infty)$ along the real axis. In the complex $t$-plane then, we have branch cuts along the positive and the negative imaginary axes.  These cuts are repeated with $2\pi$ periodicity along the real axis.  This is in contrast with the two-point function in the vacuum, which does not have these branch cuts. The appearance of the branch cuts is a  singular modification of the analytic structure. We should think of these branch cuts as appearing because of the \emph{condensation} of an infinite number of thermal poles. Indeed, in the superstar ensemble $\beta \sim  \frac{1}{N}$, so the thermal periodicity is $O(1/N)$.  Thus the thermal poles condense into branch cuts in the large $N$ limit. This should be interpreted as the effective, long-distance result -- if we had $O(N^{-1/2})$ resolution, then we could zoom in to the location of the branch cut and resolve it. The same behavior is also expected in the $\cp\cp$ two point function in a typical microstate, which is a manifestation of eigenstate thermalization in position space. It would be interesting to understand how or whether probe corrections (i.e., $O(k^2/N)$ corrections) resolve these singularities, along the lines of \cite{Faulkner:2017hll}.

\subsubsection*{ Higher order correlators and transitions}

So far we have only considered one and two point functions. One could worry about higher point functions, but  all higher point correlators satisfy large $N$ factorization, so they are suppressed relative to the two point function.
For instance, the connected three point function can be evaluated in the superstar ensemble \eqref{supens} to be  
\begin{align}
\langle t_{k_1,l_1} t_{k_2,l_2} t_{k_3,l_3} \rangle_\star = N \frac{(1-2 u_0)(1- u_0) u_0^{-(k_1+k_2+k_3)}}{1+k_1+k_2+k_3} \delta_{k_1+k_2+k_3,l_1+l_2+l_3} + O(N^0) \, .
\end{align}
This gives
\beq
\frac{\langle t_{k_1,l_1} t_{k_2,l_2} t_{k_3,l_3} \rangle_\star}{\prod_{i=1}^3\sqrt{\langle t_{k_i,l_i}t_{-k_i,-l_i}\rangle_{\star}}} \sim O(N^{-1/2}) \, .
\eeq
In our case universal behavior is only expected in the large $N$ limit which is the analog of the thermodynamic limit in our setting, so in contrast with less symmetric versions of holography, we can only expect universality to leading order in $N$. The calculation above illustrates that as $N \to \infty$, because of large $N$ factorization, it suffices to ask whether the two-point function is universal \footnote{If one starts computing directly  with Young diagrams, one can show that since traces act locally on the edge of the diagram, higher point functions of the traces look like those that are drawn from a multivariable Gaussian distribution: they are determined by the two point functions. Understanding the consequences of the Gaussian statistics is important. We are currently looking into this \cite{all:future}.}.


Corrections to the universality of code subspace correlators could also appear as non-vanishing transition elements $\langle \alpha | \mathcal{O} | \alpha'\rangle$ between different reference states (see discussion around \eqref{ETHlike}).  However, such transitions can only occur if the operator is sufficiently complex.  In the language of our Young tableau, two reference states will differ in the placement of at least $O(\sqrt{N})$ boxes.  But the $t_k$ operators add or subtract parametrically fewer boxes and so cannot produce these transitions.  More concretely, we expect that transition elements between code subspaces built around different reference states will be suppressed in powers of the entropy.

To see this, consider the average of $\langle \alpha | \cO | \alpha' \rangle$ over the micro-canonical ensemble for some simple operator in our code subspace. This quantity diagnoses the size of the off-diagonal corrections to \eqref{ETHlike}. The action of a simple operator made of $t_k$ will be to add or remove hooks from a given tableau as was explained in \cite{Berenstein:2017abm}.  This paper showed that the number of hooks of a given length scales polynomially with the number of rows in a tableau. So the action of a $t_k$ or a polynomial combination of them will take any particular tableau to a superposition of polynomially many other tableaux. The overlap above between two microstates will only be non-zero if $|\alpha\rangle$ is one of the polynomially many states reached from $|\alpha'\rangle$.  Even when the overlap is non-zero it will be polynomially suppressed because the probability spreads over the number of hooks, and so the likelihood of ending up in a particular tableau is suppressed. The dimension of the micro-canonical subspace scales like $e^N$, since the entropy of the superstar is linear in $N$. Therefore, once averaged over the ensemble, the off-diagonal elements will be suppresed by the large phase space which simple operators cannot explore 
\begin{align}
\mathbb{E}\left[ \frac{|\langle \alpha | \cO | \alpha' \rangle|^2}{\lVert \cO | \alpha \rangle \rVert \;  \lVert \cO | \alpha' \rangle \rVert} \right]
&\sim e^{-N} Poly(N) \,.
\end{align}
This shows that off-diagonal corrections to universality are suppressed in the thermodynamic (large-N) limit.
A similar logic applies to higher moments in the micro-canonical ensemble of off-diagonal correlation functions, so that this should be understood as a general suppression of off-diagonal terms and not simply a fact about the average. 

\paragraph{Summary:} We have argued that when a substantial fraction $e^{\varepsilon S}$ of the total states of a highly degenerate system such as the half-BPS superstar are probed with only simple operators, the Hilbert space naturally appears to admit a tensor factorization between the coarse ``exterior'' degrees of freedom and the fine ``interior'' degrees of freedom. This factorization is reminiscent of the arguments in \cite{Papadodimas:2012aq, Papadodimas:2013wnh}. Starting with any reference microstate, we showed that to remain within the code subspace we can apply any combination of the $t_{k,l}$ operators such that the total $k$ plus the total $l$ are much less than $\sqrt{N}$.  Since $k$ and $l$ are themselves restricted to be less than $\sqrt{N}$, there are at most $\sqrt{N} \times \sqrt{N}$ operators that we can consider.  The number of monomials built from these $t_{k,l}$ with total $k$ and $l$ less than $\sqrt{N}$ is therefore upper-bounded by $2^N = e^{N \ln{2}}$.  This is a substantial overestimate, but still shows that the code subspace which can be reached by the action of our low-energy operators (defined by preserving large $N$ factorization) is exponentially smaller than the full Hilbert space which has $e^S$ states where $S$ is the entropy.\footnote{Eq.~\eqref{SSentropy} shows that for the superstar where $N_c = N$ the entropy is much bigger than $N \ln 2$.}   Because of this and the universality of correlators in each code subspace, the total Hilbert space looks approximately factorized as ${\cal H} \sim {\cal H}_{SS} \otimes {\cal H}_{univ}$ when probed with simple, low-energy operators. This is a  realization of the general arguments presented in Sec.~\eqref{sec:genpic}.


\section{Symplectic form: Gravitational Analysis}\label{SFGrav}

\subsection{AdS/CFT in the half-BPS sector}
The 1/2-BPS states in $\cN=4$ SYM are dual to a class of asymptotically AdS$_5$ solutions in Type IIB supergravity, which involve the metric and the 5-form flux, constructed by Lin, Lunin and Maldacena (LLM) \cite{Lin:2004nb}. The metric for these LLM geometries takes the form
\beq \label{LLMmet}
g = - h^{-2} \left(dt^2 + V_i dx^i\right)^2 + h^2\left(dy^2 + dx^idx^i\right)+ye^Gd\Omega_3^2 + ye^{-G} d\tilde{\Omega}_3^2,
\eeq
where $y \in \mathbb{R}_+,\;x^i \in \mathbb{R}^2$, $d\Omega_3^2$ and $d\tilde{\Omega}_3^2$ are the standard metrics on two 3-spheres $S^3$ and $\tilde{S}_3$, and $t$ is the time coordinate.  By convention, the coordinates $(y, x^1, x^2)$ have the units of length$^2$ or area in gravitational units. The various functions appearing in this metric can all be expressed in terms of one function $z(y,x_1,x_2)$:
\beq
h^{-2} = \frac{y}{\sqrt{1/4-z^2}}, \;\; e^{2G} = \frac{1/2+z}{1/2-z}.
\eeq
\beq
y\pa_y V_i = \epsilon_{ij} \pa_j z,\;\;\;  y\left(\pa_i V_j - \pa_j V_i\right) = \epsilon_{ij} \pa_y z.
\eeq
Further, $z$ solves the following differential equation:
\beq
y\pa_y\left(\frac{1}{y}\pa_y z\right) + \pa_i\pa_i z = 0,
\eeq
with the boundary condition $\lim_{y\to 0}z(y,x_1,x_2) = z_0(x_1,x_2)$, which has the solution
\beq
z(y,x_1,x_2) = \frac{y^2}{\pi}\int d^2x'\,\frac{1}{\left(y^2+|x-x'|^2\right)^2}z_0(x_1',x_2').
\eeq
For the metric \eqref{LLMmet} to be regular in the limit $y\to 0$, the boundary condition $z_0(x^i)$ can only take on the values $\pm \frac{1}{2}$. On the regions where $z_0=+\frac{1}{2}$ (which we may choose to represent as white regions), $S^3$ shrinks smoothly as $y\to 0$, while on the regions with $z_0=-\frac{1}{2}$ (which we may choose to represent as black regions), $\tilde{S}^3$ shrinks smoothly.
Coming to the 5-form flux, the solution takes the form
\beq
F_5 = dB \wedge \mathrm{vol}_{S^3} + d\tilde{B} \wedge \mathrm{vol}_{\tilde{S}^3},
\eeq
where the one-forms $B$ and $\tilde{B}$ are given by
\beq
B_t = -\frac{1}{4}y^2e^{2G},\;\;\; \tilde{B}_t =  -\frac{1}{4}y^2e^{-2G},
\eeq
\beq
B_i = -\frac{y^2V_i}{4\left(\frac{1}{2}- z\right)} - \frac{U_i}{4} - \frac{x_1}{4} \delta_{i,2},
\eeq
\beq
\tilde{B}_i = -\frac{y^2V_i}{4\left(\frac{1}{2} + z\right)} - \frac{U_i}{4} + \frac{x_1}{4} \delta_{i,2}.
\eeq
Finally, $U_i$ satisfies
\beq \label{Uieq}
\pa_y U_i = -2y V_i.
\eeq

 The correspondence with 1/2 BPS states in $\cN=4$ SYM \cite{Lin:2004nb} proceeds by identifying the LLM plane coordinatized by $(x_1,x_2)$ with the one-particle phase space of the fermionic eigenvalues in the matrix model corresponding to the 1/2 BPS sector of SYM, and setting the fermionic phase space density to be
\beq
u(q,p) = \frac{1}{2} - z_0(q,p).
\eeq
In this correspondence, $\hbar$ in the field theory is mapped to $\ell_P$ in gravity, while $N \hbar$ corrsponds to $\ell_{AdS}$; more precisely
\beq
\hbar = 2\pi \ell_P^4,\;\; \ell_{AdS}^4 = 2N\hbar.
\eeq
With this identification, the Fermi-sea in the matrix model corresponds to $AdS_5\times S^5$ on the supergravity side; small shape perturbations on the surface of the Fermi sea correspond to gravitons moving on this background. A single column Young tableau of length scaling with $N$ can be thought of as a \emph{giant graviton}, i.e., a D3 brane wrapping an $S^3 \subset S^5$ and rotating in the other two directions on the $S^5$ with angular momentum equal to the length of the column \cite{McGreevy:2000cw,Balasubramanian:2001nh}. From the phase space density point of view, this roughly corresponds to a single white ring in the black disc (the Fermi sea) of area $\hbar$. A generic Young tableau can therefore be interpreted as a bound state of giant gravitons corresponding to its columns. 

For the states of interest to us, namely typical states in the superstar ensemble, the phase-space density consists of a large number of black and white rings (Fig.~\ref{figCG}); fixing the number of columns in the superstar ensemble is equivalent to fixing the total number of giant gravitons. If the widths of these rings were small compared with $\ell_{AdS}$ but large compared with $\ell_P$,  this would correspond to a perfectly \emph{regular}, albeit topologically complex geometry. However, as we observed earlier, the typical state will consist of densely packed rings of area $\hbar \leftrightarrow \ell^4_P$.  We could think of this as a ``Planck-scale foam" that a classical observer cannot directly measure.  Instead such an observer will only have access to the \emph{coarse-grained} phase space density which we described above, i.e., the grey disc.  Translated into gravity this corresponds to a boundary condition $z_0(x_1,x_2)$ which is not $\pm 1/2$.   Such a boundary condition does not correspond to a regular geometry -- as explained above, the only regular boundary conditions are $z_0=\pm 1/2$. Indeed, if we use any other boundary condition in constructing the LLM metric,  the resulting geometry has singular behavior in the $y\to 0$ limit. In \cite{Balasubramanian:2005mg}, it was argued that this is the origin of black hole singularities in general relativity -- the singular geometry corresponds to a coarse-grained description of a large number of underlying microstates. 

In the present paper, our focus will be on studying the gravitational phase space around these black hole-like geometries. We will see in the following section that there are surprising features which arise in the emergent phase space owing to the classical coarse-graining of the underlying space of states.

\subsection{Symplectic form around the superstar}
We want to study the gravitational phase space in Type IIB supergravity restricted to 1/2 BPS geometries around the superstar.  The authors of \cite{Grant:2005qc, Maoz:2005nk} constructed the necessary symplectic 2-form in the special cases corresponding to the vacuum and particular excited states with non-singular gravitational descriptions. However, in our case the background geometry is singular, and the calculation leads to some novel features. 

For simplicity we will consider a superstar geometry corresponding to a Young tableau state with $N$ columns (essentially $N$ giant gravitons), i.e., $\omega=1$.  This geometry corresponds to the boundary condition: 
\beq\label{SS}
z_0(x_1,x_2) = \begin{cases} 0 & \cdots\;\;  r < R \\  1/2 &\cdots \;\; r > R, \end{cases}
\eeq
where $R^2 = 4N\hbar = 2\ell_{AdS}^4$. More generally, if the number of columns is not equal to $N$, the value of $z_0$ would be a real number between $(-1/2,1/2)$ in the region $r<R$.  As discussed above, the grey disc corresponds to a singular geometry --  we should really regard it as an averaged, or coarse-grained version of regular microstates consisting of a large number of black and white regions, each of which is much smaller than the coarse-graining scale, so that the coarse-grained boundary-condition looks like a grey disc (see Fig.~\ref{figCG}).

We wish to study the phase space of geometric deformations around the superstar. For instance, we may consider position dependent ``greyscale deformations'' 
\beq
-\frac{1}{2} \leq \delta z_0(x^i) \leq \frac{1}{2}
\label{greydef}
\eeq
inside the grey disc. In order to deal with the singularity, we will consider the ``stretched'' LLM plane, namely a surface at $y = y_0$, where $y_0 \gg \ell^2_P \sim R^2_{AdS}/N^{1/2}$, and $y_0 \ll R \sim O(1)$. More precisely, we want to take the limit 
\beq\label{dsl}
\ell_P \to 0,\; y_0 \to 0,\;\mathrm{with}\;y_0/\ell^2_P \to \infty.
\eeq
We will think of the grey-scale deformations as boundary conditions at $y=y_0$. As above, any such boundary condition corresponds to a large class of microstates, which upon coarse-graining look like the specified boundary condition on the stretched LLM plane. We wish to study the phase space of these coarse-grained solutions. 
To see why this might be a reasonable thing to do, note that the evolution from $y=0$ to $y= y_0$ is implemented by the  kernel:
\beq
z(y_0,x_1,x_2 ) = \frac{1}{\pi} \int d^2x'\,\frac{y_0^2}{\left(y_0^2+ (x-x')^2\right)^2} z_0(x_1',x_2').
\eeq
This kernel has the effect of coarse graining the boundary condition over the scale $y_0$. This is because the kernel $P_{IR}(x-x')\equiv \frac{1}{\pi} \frac{y_0^2}{\left(y_0^2+ (x-x')^2\right)^2}$ is given in momentum space by
\beq \label{GravProj}
P_{IR}(k) \sim |k| y_0 K_1(|k| y_0),
\eeq
which is essentially a ``smooth filter'' cutting off the UV modes, namely $P_{IR}(k) \to 1 $ for $|k|y_0 << 1$ and $P_{IR}(k) \to 0$ for $|k| y_0 >>1$. In Sec.~\ref{SFCFT}, when we study the symplectic form from the CFT point of view, we will use this type of a cutoff on UV modes in order to obtain a coarse-grained description, which we will then compare with the gravity result.

\subsubsection*{Stretching the LLM plane}
Recall from above that all terms in the LLM solution are controlled by a single function $z$ which solves
\beq
y\pa_y\left(\frac{1}{y}\pa_y z\right) + \pa_i\pa_i z = 0.
\eeq
This has  solutions
\beq
z(y, k^i) = y \left(c_1 K_1(\kappa y) + c_2 I_1(\kappa y)\right).
\eeq
where $\kappa = \sqrt{k^2}$. The solution $I_1$ diverges for large $y$, so we discard it on physical grounds. On the other hand, while $K_1(\kappa y)$ diverges logarithmically for $y \to 0$, the combination $y K_1(\kappa y)$ remains finite. In order to determine $c_1$, we impose  boundary conditions at the stretched plane $y = y_0$: 
\beq
z(y_0, k) = z_0(k).
\eeq
This gives the final solution in momentum space
\beq
z(y,k) =  \frac{y K_1(\kappa y) }{y_0 K_1( \kappa y_0) } z_0(k).
\eeq
In position space, this translates to
\beq\label{ker}
z(y,x^i) = \int d^2x' \cA(y | x,x') z_0(x'),\;\; \cA(y | x,x') = \int \frac{d^2k}{(2\pi)^2} e^{i k\cdot (x- x')}\frac{y K_1(\kappa y) }{y_0 K_1( \kappa y_0) }.
\eeq
In the limit $y_0 \to 0$, the integral kernel simplifies greatly and we find\footnote{Using $\lim_{y_0\to 0}y_0K_1(\kappa y_0) = \kappa^{-1}$.}
\beq
\lim_{y_0\to 0} \cA(y | x,x') =
\frac{y^2}{\pi}\frac{1}{\left(y^2+|x-x'|^2\right)^2},
\eeq
which is the familiar LLM kernel described above.  
Similarly, the vector field $V_i$ 
with boundary conditions at the stretched surface takes the form
\beq
V_i(y,x^i) = -\epsilon_{ij}\pa_j\int d^2x' \cB(y | x,x') z_0(x'),\;\; \cB(y | x,x') = \int \frac{d^2k}{(2\pi)^2} e^{i k\cdot (x- x')}\frac{ K_0(\kappa y) }{\kappa y_0 K_1( \kappa y_0) }.
\eeq

Now coming to the 5-form flux, the solution takes the form
\beq
F_5 = dB \wedge \mathrm{vol}_{S^3} + d\tilde{B} \wedge \mathrm{vol}_{\tilde{S}^3}.
\eeq
The one-forms $B$ and $\tilde{B}$ are given by
\beq
B_t = -\frac{1}{4}y^2e^{2G},\;\;\; \tilde{B}_t =  -\frac{1}{4}y^2e^{-2G},
\eeq
\beq
B_i = -\frac{y^2V_i}{4\left(\frac{1}{2}- z\right)} - \frac{U_i}{4} - \frac{x_1}{4} \delta_{i,2},
\eeq
\beq
\tilde{B}_i = -\frac{y^2V_i}{4\left(\frac{1}{2} + z\right)} - \frac{U_i}{4} + \frac{x_1}{4} \delta_{i,2}.
\eeq
Finally the equation for $U_i$ \eqref{Uieq}, can be solved to get
\beq
U_i = -2\epsilon_{ij}\pa_j \int d^2x'\,\cC(y|x,x')z_0(x'), \;\;\; \cC(y | x,x') = \int \frac{d^2k}{(2\pi)^2} e^{i k\cdot (x- x')}\frac{y K_1(\kappa y) }{k^2 y_0 K_1( \kappa y_0) }.
\eeq
Care should be taken while using the integral definition of the kernel $\cC$. For example, in the $y_0 \to 0$ limit,  the $k$ integral naively diverges, but the kernel $\pa_i \cC$ is well-defined. Finally, we can conveniently write the solutions for $z, V_i$ and $U_i$ using the convolution notation:
\beq
z = \cA*z_0, \;\;V_i = -\epsilon_{ij}\pa_j \cB*z_0,\;\;U_i = -\epsilon_{ij}\pa_j \cC*z_0.
\eeq

\subsubsection*{Symplectic form}
\newcommand{\bs}{\boldsymbol}
\newcommand{\bd}{\boldsymbol{\delta}}
The symplectic 2-form for type IIB supergravity in the LLM sector was worked out in \cite{Grant:2005qc, Maoz:2005nk}:
\beq\label{SF}
\bs{\Omega} = \int_{y=y_0} d^2x\;\bs\omega,\;\;\;\; \bs\omega = -\epsilon_{ij} \bd V_i \bd\left(\alpha V_j\right) - \epsilon_{ij} \bd a_i \bd b_j+ 8\left( \bd \lambda \bd \tilde{F}_{12} -\bd\tilde{ \lambda} \bd F_{12}\right).
\eeq 
where
\beq
\alpha = -\frac{y^4 z\left(\frac{1}{4} + z^2\right)}{\left(\frac{1}{4} - z^2\right)^2},
\eeq
 \beq
 a_i = \frac{y^4 V_i}{2\left(\frac{1}{4} - z^2\right)}+ U_i,\;\;\; b_i = \frac{y^4 z V_i}{\left(\frac{1}{4} - z^2\right)}.
\eeq
Here we use bold symbols to denote differential forms in field space, so for instance $\bd$ is the exterior derivative in field space. Note that $\bs{\Omega}$ has been expressed here as an integral over the {\it stretched} LLM plane. Further, $\bd \lambda$ and $\bd\tilde{\lambda}$ are pure gauge modes corresponding to the gauge field $B$ and $\tilde{B}$ (which appear in the ansatz for the 4-form gauge field). For any given microscopic LLM solution, these modes are partly determined by the requirement that $\bd B^{reg.}_i = \bd B_i + \pa_i \bd \lambda$ vanishes in the limit $y \to 0$:
\beqn
\pa_i \bd\lambda &=& - \bd B_i \;\;\;\; \cdots \;\;\;(z= -1/2),\nonumber\\
 \pa_i \bd\tilde{\lambda } &=& - \bd\tilde{ B}_i\;\;\;\;\cdots \;\;\;(z=+1/2)
\eeqn
on the appropriate regions of the LLM plane \cite{Maoz:2005nk}, while on the remaining regions these modes drop out of $\bs{\Omega}$. This is a regularity condition on the deformation $\bd B^{reg.}_i$ which ensures that the geometry smoothly caps off as $y\to 0$ (see the Appendix of \cite{Maoz:2005nk} for a more detailed explanation). Crucially, note that the above fixing of the gauge-modes depends sensitively on the background microstate. 

As a consequence of these boundary conditions, $\bd\lambda, \bd\tilde{\lambda}$ do not ultimately appear in the symplectic form -- rather, they are determined in terms of $\bd z_0$. However, in the present case since we are imposing boundary conditions at $y=y_0$, we do not have any such regularity conditions to fix the pure gauge modes $(\bd \lambda, \bd \tilde{\lambda})$. Consequently, these modes remain \emph{dynamical} because they cannot be solved for in terms of other quantities on phase space. An analogous situation occurs in gauge theories, where consistently restricting to a subregion leads to the appearance of \emph{soft modes} on the boundary.  These soft modes are erstwhile gauge degrees of freedom that become physical on the boundary of a subregion.  In our case, the subregion in question lies in the range $y > y_0$.  In the double-scaling limit relevant to us \eqref{dsl}, i.e., $\ell_P \to 0, y_0 \to 0$ with $y_0/\ell_P^2 \to \infty$, our soft modes do not decouple.

We may interpret our soft modes as parametrizing our ignorance of the microscopic state.     If we had access to the detailed microstate geometry, then imposing regularity would have fixed $(\bd \lambda, \bd \tilde{\lambda})$ in terms of $\bd z_0$.  However, a classical observer cannot resolve geometry at this scale -- indeed, in a quantum theory we should not even talk about geometry at this scale.   The resulting ambiguity in the background microstate is then reflected in the presence of a new mode which can be  interpreted as shifting between microscopic configurations that give the same macroscopic geometry -- this will be further clarified by the CFT discussion in section 5.   Our double scaling limit $\ell_P \to 0, y_0 \to 0$ with $y_0/\ell_P^2 \to \infty$ is chosen to make this clear by washing out the region of spacetime, which does not have a classical interpretation.   In this limit we recover the superstar geometry as a universal description of many microstates, but this comes at the cost of a soft mode on phase space that encodes the underlying ambiguity.


In order to finish the computation of the symplectic form, we must obtain $\bd V_i$ and $\bd U_i$ in the $y \to y_0$ limit.  First,
\beq
\bd V_i(y_0,x ) = -\epsilon_{ij} \pa_j \cB(y_0)*\bd z_0 = -\epsilon_{ij}  \cB(y_0)*\pa_j\bd z_0 .
\eeq
where the kernel $\cB(y_0)$ is given by:
\beq
\cB(y_0)  =  \int \frac{d^2k}{(2\pi)^2} e^{i k\cdot (x- x')}\frac{ K_0(\kappa y_0) }{\kappa y_0 K_1( \kappa y_0) }.
\eeq
On the other hand, we have a closed form expression for $\bd U_i$
\beq \label{bdu}
\bd U_i(y_0,x) = -2\epsilon_{ij} \pa_j \cC(y_0)*\bd z_0 = \frac{1}{\pi} \epsilon_{ij} \int d^2x'\; \frac{(x-x')_j}{|x-x'|^2}\bd z_0(x').
\eeq
Here $\bd z_0$ is the grey-scale deformation \eqref{greydef} around the superstar boundary condition on the stretched LLM plane. 


As discussed above, the symplectic form involves an integral over the LLM plane at $y = y_0$.  We can divide this plane into three regions: (i) the \emph{interior} $r < (R - \varepsilon)$, (ii) the \emph{thickened boundary} $(R-\varepsilon) \leq r \leq (R + \varepsilon)$, and (iii) the \emph{exterior} $r > (R + \varepsilon)$ (where $\varepsilon \sim y_0$ is the thickness of the boundary).  In Appendix~\ref{appA}, we show that the integral over the thickened boundary and the exterior vanish.  It remains to compute the integral over the interior.

 In the interior region, $z_0= 0$, and so only the last term in the symplectic form \eqref{SF} contributes as $y_0 \to 0 $ because the other terms vanish polynomially in the limit:
\beq
\bs{\Omega}_{int.} = 8\int d^2x\,\left( \bd \lambda \bd \tilde{F}_{12} -\bd\tilde{ \lambda} \bd F_{12}\right).
\eeq 
Now, we may use
\beq
\bd B_i = -\bd\left[\frac{y^2V_i}{4\left(\frac{1}{2}- z\right)}\right] - \frac{\bd U_i}{4} 
\eeq
and similarly for $\bd \tilde{B}_i$. The first term once again drops out in the $y_0 \to 0$ limit, and so
\beq
\bd B_i = -\frac{1}{4} \bd U_i \;\; \Rightarrow \;\; \bd F_{12} = -\frac{1}{4} \epsilon^{ij} \pa_{i}\bd U_j. 
\eeq
From equation \eqref{bdu}, we have 
\beq
\epsilon^{ij} \pa_{i}\bd U_j = \frac{1}{\pi} \epsilon^{ij}\epsilon_{jk}\pa_i \int d^2x'\; \frac{(x-x')_k}{|x-x'|^2}\bd z_0(x') = 2\bd z_0(x).
\eeq
This leads to
\beq
\bs{\Omega}_{int.}= 4\int_{r<R - \varepsilon} d^2x\,\left( \bd\tilde{ \lambda}(x)-\bd \lambda(x) \right) \bd z_0(x).
\eeq 
It is conveninent to introduce the notation $\pi_{grav} = 4(\tilde{\lambda}-\lambda)$, and $u_{grav} = z_0 + \frac{1}{2}$, in terms of which we have
\beq
\bs{\Omega}= \int_{r<R} d^2x\,\bd \pi_{grav}(x) \bd u_{grav}(x).
\eeq
If we had carried out this analysis with a smooth boundary condition on the LLM plane, the soft mode $\pi_{grav}$ would have been fixed and non-dynamical.   However, in the classical geometry which coarse grains over smooth microstates through our double scaling limit, we see that $\pi_{grav}$ is dynamical (i.e., remains in the symplectic form).   In the next section, we will see how this symplectic form emerges from a dual CFT point of view, and in particular how to interpret $\bd\pi_{grav}$ from a microscopic perspective.

\section{Symplectic form: CFT analysis}\label{SFCFT}
In the previous section, we derived the symplectic form for greyscale fluctuations from the gravity point of view. We observed the emergence of a new soft mode in this description. Here we wish to give a CFT description of this mode -- we will see that it naturally appears as a consequence of coarse graining over a scale $y_0$ that is much larger than the Planck scale.  The radial cutoff we imposed on the gravity side corresponds to this scale because initial data on the LLM plane is effectively coarse-grained as it is transported to the stretched surface.

\subsection{Coherent states and coadjoint orbits}
Firstly, we must understand how to extract a classical phase space and its corresponding symplectic form from the quantum Hilbert space of the CFT, which in the half-BPS case can be expressed in terms of free fermions (as explained in section 2). This can be accomplished by using \emph{coherent states} of the density operator $\hat{u}$. As we will explain below, these coherent states allow us to identify a phase space for the greyscale fluctuations $\delta u$ defined in equation \eqref{greydef}, in the classical limit $\hbar \to 0$. The symplectic form on this phase space can be constructed by using the Kirillov-Kostant co-adjoint orbit method (see \cite{Yaffe:1981vf, Witten:1987ty} for a pedagogical introduction).\footnote{The authors of reference \cite{Dhar:1993jc} perform a closely related analysis, but in the context of regular LLM geometries (i.e., not singular solutions like the superstar). They write the result in the form of an action with ``two times'', but it is possible to read off the symplectic form from their action. Further this symplectic form (read off from \cite{Dhar:1993jc}) appears to be slightly different from our version, but it is an algebraic exercise to check that these two versions are equivalent. We thank Gautam Mandal for explaining this to us.}  

To this end, consider a class of coherent  excitations of a background state $\psi_0$: 
\beq\label{coadorb}
|\psi\rangle = e^{i\hat{T}_{\pi}} |\psi_0\rangle,\;\;\;\hat{T}_{\pi} = \int \frac{dp'dq'}{2\pi\hbar} \pi(q',p')\hat{u}(q',p').
\eeq
In the infinitesimal case this reads as
\beq \label{infCO}
\delta|\psi\rangle = i\int \frac{dp'dq'}{2\pi\hbar} \delta \pi(q',p')\hat{u}(q',p') |\psi_0\rangle. 
\eeq
This defines the \emph{coadjoint orbit} of $\psi_0$. As discussed above in Sec.~\ref{LEC}, with these definitions,
\beq\label{du-app}
\delta u = \epsilon^{ij}\pa_i \delta \pi \pa_j u = \left\{\delta \pi, u \right\}_{PB}.
\eeq
Although we defined the state deformation in terms of $\delta \pi$ initially, we can invert equation \eqref{du-app} to solve for $\delta \pi$ in terms of $\delta u$ on a given coadjoint orbit, and so we can equivalently label the state deformations by $\delta u$. 

Now, the standard symplectic form on coadjoint orbits, in our notation, is given by \cite{Yaffe:1981vf, Witten:1987ty}
\beq
\bs{\Omega}(u;\delta_1u,\delta_2u) = \hbar\left\langle \psi_0 \left| \left[\hat{T}_{\delta_1\pi}, \hat{T}_{\delta_2\pi}\right] \right| \psi_0\right\rangle,
\eeq
where $\delta_{1,2}\pi$ are related to $\delta_{1,2}u$ by equation \eqref{du-app}. Finally, using the commutation relations \eqref{CR0} (with $\hbar \to 0$) and simplifying gives the  result:
\beq\label{sfnew}
\bs{\Omega}(u; \delta_1u, \delta_2u) = \int \frac{dpdq}{2\pi}\, u \left\{\delta_1 \pi, \delta_2\pi\right\}_{PB}.
\eeq 
where the Poisson bracket is defined by $\left\{A, B\right\}_{PB} = \epsilon^{ij}\pa_i A \pa_j B$ (where $\pa_i = (\pa_p, \pa_q)$ etc., and recall that the one-particle phase space coordinates $(p,q)$ are to be identified with the coordinates on the LLM plane from the gravity point of view). The quantity $\delta \pi$ is defined implicitly in terms of $\delta u$ as in \eqref{du-app}.
Integration by parts gives an identity
\beq
\int dp \, dq \, A\left\{B,C\right\} = \int dp \, dq \, B\left\{C, A\right\},
\eeq 
which we can use to rewrite the symplectic form as
\beqn \label{SF1}
\bs{\Omega}(u; \delta_1u, \delta_2u) &=&\frac{1}{2} \int \frac{dpdq}{2\pi}\; \left(\delta_1 \pi \left\{ \delta_2\pi , u\right\}_{PB}-\delta_2 \pi \left\{ \delta_1\pi , u\right\}_{PB}\right)\nonumber\\
&=&\frac{1}{2} \int \frac{dpdq}{2\pi}\;\left( \delta_1\pi \, \delta_2 u - \delta_2\pi \, \delta_1u\right).
\eeqn
In the second line we have used the definition \eqref{du-app}. We see that the field we denoted by $\delta \pi$ above is indeed precisely the canonical momentum.

Note that the above symplectic form can also be written as $\hbar$ times the Berry-curvature (see \cite{Belin:2018fxe} for related discussion in AdS/CFT): 
\beq
\bs{\Omega}(u_0; \delta_1u, \delta_2u) = \hbar F_{Berry}\equiv i\hbar\Big(\langle \delta_1 \psi | \delta_2\psi \rangle - \langle \delta_2\psi | \delta_1\psi\rangle\Big). 
\eeq
From the geometric quantization of a classical phase space it is familiar that the curvature of the prequantum line-bundle is $1/\hbar$ times the classical symplectic form. Here, we are going in the opposite direction, by extracting the classical symplectic form from the quantum Berry curvature. It is not obvious that one can always recover the classical symplectic form from the quantum Hilbert space in this way. However, it does work in cases where the Hilbert space is obtained by suitably quantizing a coadjoint orbit.

\begin{figure}[!h]
\centering
\includegraphics[height=4cm]{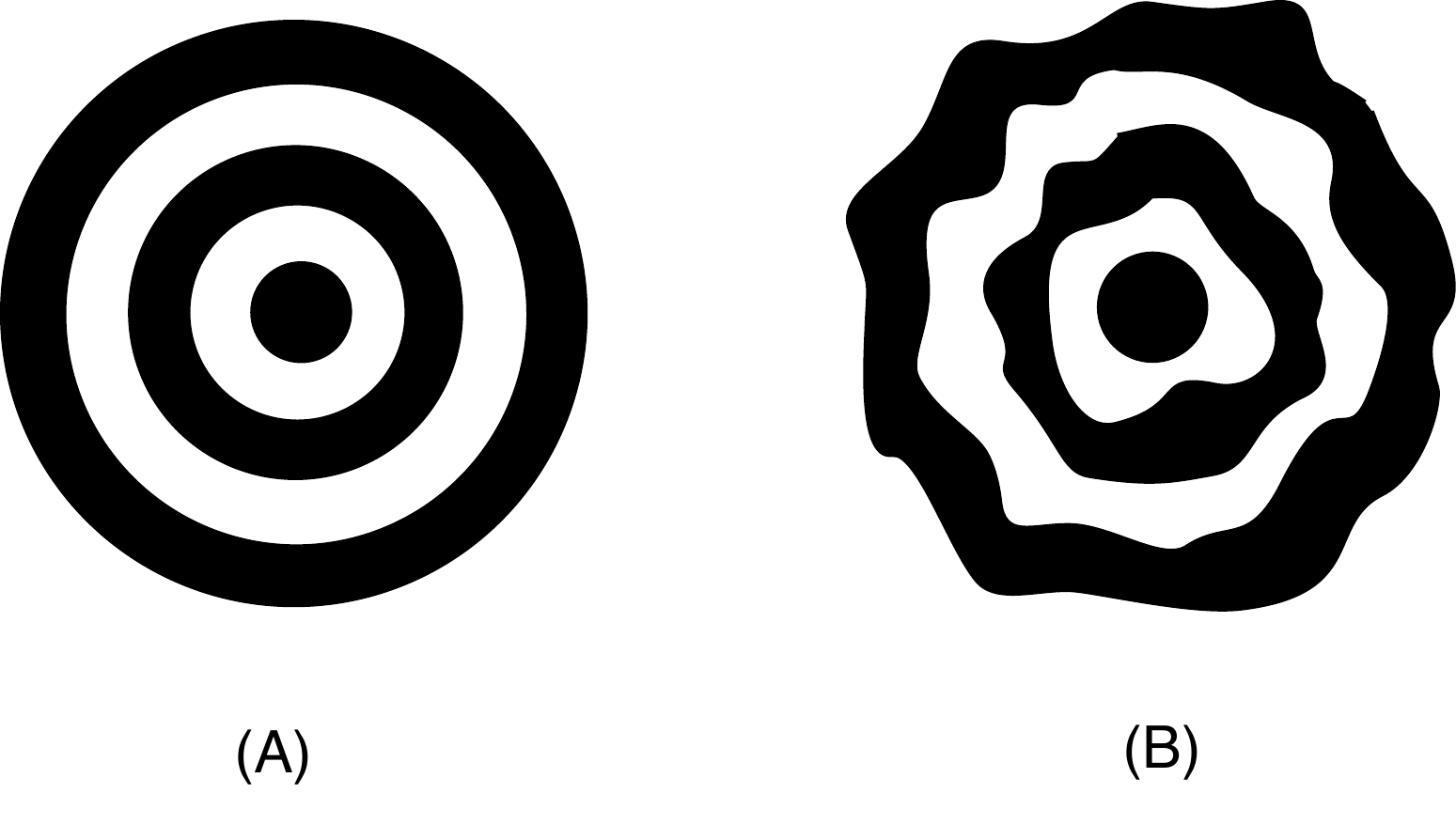}
\caption{\small{\textsf{(A) The background configuration consisting of annuli or rings. (B) A sample classical deformation around this configuration.\label{fig:rings}}}}
\end{figure}

It is easy to demonstrate that 
\eqref{sfnew}  correctly reproduces the results of \cite{Maoz:2005nk} for the symplectic form in the specific case where the background density is given by concentric rings (Fig.~\ref{fig:rings}).  In fact, this is essentially the background that will be of interest for us in the context of the superstar, albeit with a large number of rings. So, consider the background density:
\beq \label{rings}
u(r) = \sum_i (-1)^i \Theta(r_i - r),
\eeq
where even and odd $i$'s correspond to edges and anti-edges.  In this case,  $u$ must be 0 or 1 everywhere microscopically, and thus deformations can only occur at the edges of the rings (i.e., deformations of the shape of the edges) if we want to preserve the topology of the bands.   Correspondingly, we must take the deformation $\delta u$ to be of the form
\beq \label{du0}
\delta u(r,\theta) = \sum_i (-1)^i \delta r_i(\theta) \delta(r_i - r)= \sum_i (-1)^i \delta \gamma_i(\theta) \frac{1}{r_i}\delta(r_i - r).
\eeq
where $\delta \gamma = \delta\left(\frac{1}{2} r^2\right)$.   Next, in order to obtain $\delta \pi$, we must solve \eqref{du-app}, and this gives
\beq
\sum_i (-1)^i \delta \gamma_i(\theta) \frac{1}{r_i}\delta(r_i - r) = \sum_i (-1)^i \pa_{\theta}\delta \pi(r_i,\theta) \frac{1}{r_i}\delta(r_i - r),
\eeq
where we are using the convention $\epsilon^{r\theta} = \frac{1}{r}$ for the Levi-Civita symbols. Comparing the two sides, we see that the $\delta \pi$s at the boundaries of the rings are determined as
\beq
\pa_{\theta} \delta \pi(r_i,\theta) = \delta \gamma_i(\theta)\;\;\Rightarrow \;\; \delta \pi(r_i,\theta) = \frac{1}{2}\int_0^{2\pi}d\theta' \mathrm{Sign}(\theta - \theta') \delta \gamma_i(\theta').
 \eeq
Returning to the symplectic form, we get (switching to form notation in field space for simplicity) 
\beq
\bs{\Omega} =  \frac{1}{2}\int \frac{dpdq}{2\pi}\; \bs\delta \pi \bs \delta u =  \sum_i \frac{(-1)^i }{8\pi}\int d\theta d\theta'\; \bs\delta \gamma_i (\theta)\mathrm{Sign}(\theta- \theta') \bs\delta\gamma_i(\theta').
\eeq
This is the result of Maoz and Rychkov \cite{Maoz:2005nk}. Note that although the momentum is only determined at the boundaries of the droplet (and generally undermined away from the boundaries), the symplectic form is completely well-defined, as it is localized on the boundaries. 
As in the previous section, we are using $\bd$ to denote antisymmetrized variations. 

\subsection{Superstar}
We wish to apply the above phase space analysis to the CFT state corresponding to the superstar, but there is a subtlety.  The typical microstates which are described universally by the superstar geometry appear in phase space as a dense collection of concentric black and white rings. The separations between these rings are at the $\hbar$ scale (corresponding to the Planck scale in the dual gravity description).  Of course, a classical description of such configurations is not possible.   However, we can consider a coarse-graining scale $y_0$ that is much larger than $\hbar^{1/2}$.  In this case there will be many configurations with microstructure at some scale $\epsilon$ much smaller than the coarse-graining scale $y_0$, but nevertheless much larger than $\hbar^{1/2}$.   The classical symplectic form analysis can be applied to such configurations and can be compared with the gravitational analysis that we described in the previous section. In the context of  section $3$, the setup we are considering is a set of $|\alpha\rangle$-microstates which have microstructure at the scale $\epsilon \gg \hbar^{1/2}$. In this way, we are considering classical deformations in the universal code subspace as (a projection of) those of \eqref{coadorb}. Within this universal code subspace we we will make a distinction between UV and IR by denoting whether fluctuations are bigger or smaller than the scale set by $y_0$. We expect that the lessons learned in this controlled example will teach us about the more interesting regime where the black and white rings are separated at the $\hbar$ scale. In the context of section $3$, this subspace's dimension is $O(N^0)$. 

Following this philosophy, we  work with  a background superstar microstate configuration made up of rings, such as in equation \eqref{rings}, where the spacing between two rings is $r_{i+1}- r_i = \epsilon$. Here $\epsilon$ is a  scale which we may take to be such that  $\hbar^{1/2} \ll \epsilon \ll y_0 \ll (N \hbar)^{1/2}$, where $y_0$ is our coarse-graining scale. Without loss of generality, we can take the variations of the phase space density around this background to be of the form
\beq \label{du1}
\delta u(r,\theta) = \epsilon \sum_{i} r_i \delta u_i(\theta) \frac{1}{r_i}\delta(r-r_i) \, .
\ee
Said differently, this just defines $\delta u_i(\theta)$. Now, from the above discussion, we can explicitly construct the conjugate momentum $\delta \pi$. Of course, $\delta \pi$ is only determined at the boundaries of the rings.  But as we showed above, we can extend $\delta \pi$ in an arbitrary but smooth manner away from these boundaries, because the symplectic form will only depend on the boundary contribution anyway. To define such an extension, let $H_{\epsilon}(x-x_0)$ be some smooth function which takes the value 1 for $|x-x_0| < \epsilon/2$, and smoothly drops to zero outside. We can then write 
\beq
\delta \pi(r,\theta) = \frac{1}{2}\int d\theta' \mathrm{Sign}(\theta- \theta')\sum_i \epsilon r_i (-1)^i \delta u(r_i, \theta') H_{\epsilon}(r-r_i).
\eeq
We can now write the symplectic form in terms of $\delta u$ and $\delta \pi$ as described in \eqref{SF1}:
\beqn\label{SFbCT}
\bs{\Omega} &=& -\frac{1}{4\pi}\int dpdq \bd u(p,q) \bd \pi(p,q)\nonumber\\
&=& -\frac{1}{4\pi}\int dr \int d\theta d\theta'  \bd u(r,\theta)\mathrm{Sign}(\theta-\theta') \sum_i \epsilon r_i (-1)^i \bd u(r_i, \theta') H_{\epsilon}(r-r_i).
\eeqn

\begin{figure}[!h]
\centering
\includegraphics[height=4cm]{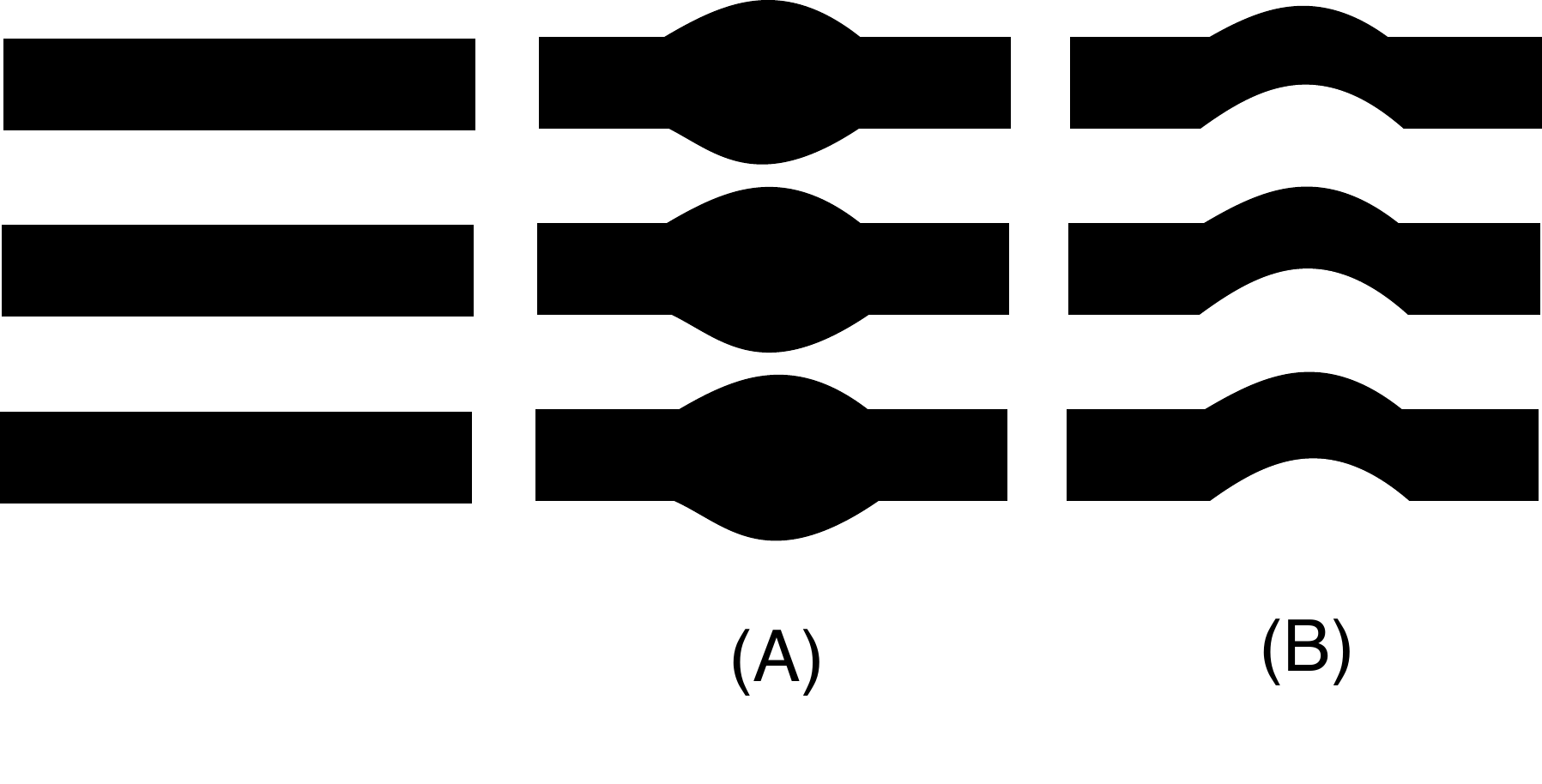}
\caption{\small{\textsf{The $A$ and $B$ modes respectively correspond to local squeezing/dilations and local translations. }}}
\end{figure}

So far we have obtained the symplectic form in terms of the phase space variable $\bd u$, which corresponds to microscopic deformations. Next, we wish to extract the symplectic form for \emph{coarse-grained} phase space variables, which we can compare with the gravitational results of the previous section. We can accomplish this by writing $\bd u$ in the form
\beq \label{CT}
 \bd u(r,\theta)= \bd A(r,\theta) + \frac{e^{-i\pi r/\epsilon}}{\epsilon} \bd B(r,\theta) ,
\eeq
where both $\bd A$ and $\bd B$ are slowly varying functions of $r$ and $\theta$ (i.e., they contain Fourier modes with wavelengths much larger than $\epsilon$). Notice, however, that because of the factor of $e^{-i\pi r/\epsilon}$, the second term in equation \eqref{CT} as a whole has rapid oscillations and is therefore in the UV part of $\bd u$. (Here by UV or IR we are referring to the coarse-graining scale $y_0$.) Equivalently,  the emergence of these two long-wavelength modes can be thought as coming from considering the excitations of edge and anti-edge independently. By comparing equations \eqref{du0} and \eqref{du1} -- which shows $\delta r_i(\theta) = (-1)^i \epsilon \delta u_i(\theta)$ -- we see that $\bd A$ is a \emph{squeezing} mode, which locally squeezes the white rings and dilates the black rings or vice versa, while $\bd B$ is a \emph{translation} mode which uniformly shifts all the rings upwards or downwards.  Therefore, it is clear that upon coarse-graining, $\bd A$ will change the IR, effective phase space density, while $\bd B$ will not. One might worry about the $\frac{1}{\epsilon}$ in the second term in \eqref{CT}, but since $\delta r_i \propto \epsilon \, \delta u_i$ and further since $B$ is a translation mode, there is no conflict with the microscopic rings (such as collisions etc.).\footnote{In fact, the condition that $B$ is slowly varying is precisely equivalent to the statement that there are no microscopic collisions between different rings. On the other hand, it is important that the squeezing mode $A$ does not scale with $\frac{1}{\epsilon}$, as this would lead to collisions of rings.} 

We may regard equation \eqref{CT} as a \emph{canonical transformation} on phase space, i.e., a change of coordinates from the microscopic $u$ to the emergent classical variables $(A, B)$ \footnote{Technically speaking, \eqref{CT} defines a section of phase space characterized by two slowly varying functions and in this subsection, the natural position and momenta are these two modes. }. We can obtain the symplectic form in terms of these new coordinates by substituting equation \eqref{CT} into \eqref{SFbCT}:
\beqn
\bs{\Omega}&=&- \frac{1}{4\pi}\int dr \int d\theta d\theta'\left[\bd A(r,\theta) + \frac{e^{-i\pi r/\epsilon}}{\epsilon} \bd B(r,\theta)\right]\nonumber\\
&\times & \mathrm{Sign}(\theta-\theta') \sum_i \epsilon r_i e^{i\pi r/\epsilon} \left[\bd A(r_i,\theta') + \frac{e^{-i\pi r_i/\epsilon}}{\epsilon} \bd B(r_i,\theta')\right] H_{\epsilon}(r-r_i).
\eeqn
The $\bd A \bd A$ and $\bd B \bd B$ terms come multiplied with the highly oscillating factor $e^{\pm i \pi r /\epsilon}$. Since both $\bd A$ and $\bd B$ are slowly varying modes, and $H_{\epsilon}$ only has support within an $\epsilon$-neighbourhood $r=r_i$, we find that these $\bd A \bd A$ and $\bd B \bd B$ terms drop out in the limit $\epsilon \to 0$. On the other hand, in the cross terms $\bd A \bd B$, these oscillating factors cancel and we obtain in the $\epsilon \to 0$ limit\footnote{Here, we have used $\lim_{\epsilon \to 0} \frac{1}{\epsilon} H_{\epsilon}(x) = \delta(x)$.}
\beq
\bs{\Omega} = -\frac{1}{2\pi}\int dr\, r\int d\theta d\theta' \bd{A}(r,\theta) \mathrm{Sign}(\theta-\theta') \bd B (r,\theta').
\eeq

Crucially, we have now obtained the symplectic form in terms of IR effective modes. 
If we consider the IR part of $\bd u$: 
$$\bd u_{IR}(x) =  \int d^2y P_{IR}(x,y) \bd u(y), \;\;\; P_{IR}(x,y) = \int_{|k|y_0 \leq 1} \frac{d^2k}{(2\pi)^2} e^{ik(x-y)},$$
then we get
\beq
\bd u_{IR}(x) = \int d^2y P_{IR}(x,y)\left[\bd A(y) + \frac{e^{-i\pi r/\epsilon}}{\epsilon}\bd B\right]\sim \bd A(x),
\eeq
where the second term drops out because of the highly oscillatory factor.\footnote{For the same reason, the energy of $\delta u$ is dominated by $\delta A$, while $\delta B$ will only make a sub-leading in $\epsilon$ contribution; so we may call $\delta B$ a soft mode.} This prescription for coarse-graining is essentially equivalent to the coarse-graining that appears naturally in gravity -- recall that in our gravity calculation, we placed a cutoff surface at the stretched horizon $y= y_0$. But as explained around equation \eqref{GravProj}, in gravity this corresponds to a smooth momentum space cutoff $\Lambda \sim 1/y_0$ in the LLM plane. Here we have taken a sharp cutoff for convenience, but do not expect this difference to be important. At any rate, we may now write the symplectic form as
\beqn
\bs{\Omega} &=& \int dp dq\; \bd{\pi}_{IR}(x) \bd u_{IR}(x),\nonumber\\
\bd \pi_{IR}(r,\theta) &\equiv & \frac{1}{2\pi}\int d\theta' \mathrm{Sign}(\theta-\theta') \bd B(r, \theta') 
\eeqn
We see that $\delta \pi_{IR}$ is a new emergent IR mode, built out of the UV modes in $\delta u$. Of course, ultimately it depends on $\delta u$, but we extracted it by a canonical transformation out of the UV part of $\delta u$, and as such it should be treated as an independent mode in the IR. Note that, at the level of the Hilbert space, the IR label \emph{does not} denote projection into a smaller subspace, which is the usual way that we think about coarse graining. This label reflects that the two classical degrees of freedom $\delta A, \delta B$ have a natural interpretation in terms of independent pieces of $\delta u$ (or linear combinations of deformations of the edges / anti-edges), but of course they ultimately come from the same $\hat{u}$ operator. It may seem confusing that the conjugate momentum in the IR is related to a UV piece of the phase space density.  This sort of UV-IR mixing would not happen if we were working in perturbation around the vacuum.  But because we are examining fluctuations around a highly structured state, it can be that the emergent IR modes actually correspond to structural perturbations at the fine scale of the background.

Comparing with the gravitational analysis from the previous section, we now see that the pure gauge mode in gravity $\bd \pi_{grav}= 4(\bd \tilde{\lambda} - \bd\lambda)$ should be identified with CFT quantity $\bd \pi_{IR}$ in this section. In this way, the emergent pure gauge mode at the stretched horizon encodes certain microscopic degrees of freedom of the black hole in the form of an effective gravitational degree of freedom. This may be the beginning of an explanation for how gravity encodes information about microstates in terms of geometric quantities at the horizon. 

Above, we constructed the coarse-grained symplectic form  in the CFT, and showed the emergence of a new effective mode in the IR built out of the UV modes in the phase space density.  In this analysis we considered {\it localized} perturbations of phase space that correspond on the AdS side to graviton fluctuations.   But what about large deformations, e.g. changing the radii or widths of the rings in phase space?  In gravity these are large perturbations that globally move the locations and sizes of  spheres in the geometry.  After coarse-graining a typical state, such perturbations will appear as angularly-symmetric greyscale deformations (as opposed to local perturbations).  Appendix~\ref{applargedef} repeats our analysis in this setting and again shows the appearance of an emergent IR mode built from UV data.

\section{Discussion}
The modern language for understanding the universality of physics around different black hole microstates is to say that there is a ``code subspace'' of low-energy excitations around any typical microstate, and that these code subspaces are isomorphic but orthogonal to each other.   In this paper, we provided an explicit construction of such code subspaces for the superstar, which is an incipient black hole.  To do this,  we imagined a subspace of reference microstates of size $e^{\varepsilon S}$ with $\varepsilon <1$ where $S$ is the microcanonical entropy.  This is an exponentially large number of states, but is still much smaller than the total number of superstar states which is $e^S$.   Around each reference state there is a code subspace, and we argued that the code subspaces are isomorphic to each other.  Thus the union of these states can be written as a code subspace times the set of reference states ${\cal H}_\alpha$.  This is only possible because we chose ${\cal H}_\alpha$ to be sufficiently sparse so that the states are all orthogonal to each other.  Otherwise, the code subspaces around different reference states would start to overlap and the factorization would break down. Similar subspaces have recently been studied in \cite{Hayden}, where the authors argued that, in AdS/CFT, given a boundary subregion bigger than half of the size of the system, the largest subspace that can be  described in low energy effective field theory independently of the underlying state has dimension $\log dim H_R= \varepsilon S$ which can be determined geometrically. In our discussion, $\varepsilon$ was arbitrary, but, in a similar way, it had to be $<1$ for a state independent low energy effective description to be accurate. It is possible that a discussion similar to that of \cite{Hayden} can give an information theoretic interpretation to the physical limitations of working with such subspace and could put bounds on $\varepsilon$ given the constraints of a putative observer.  More concretely, these large subspaces are important when considering time evolution. For strongly interacting systems, where the energy spacing is $O(e^{-S})$, if we sample a random state in the microcanonical window $|\psi \rangle$ and we study its time evolution, we will not be able to resolve its energy levels until we reach $t \sim e^{S}$. This means that for a given $t_0$, we can think of the span of $|\psi(t<t_0)\rangle$ as a subspace of dimension $O(t_0)$, so these large subspaces are all that a low energy observer has access to if she can only make long wavelength observations for a finite amount of time. In our set-up, because the spectrum is highly degenerate, the situation is not quite the same, but a similar story should still be applicable if we were to consider evolution with a non-gauge invariant Hamiltonian, which evolves the single energy eigenstates independently.

There may be a more general way to formulate the code subspace factorization in terms of a highly redundant description of the Hilbert space. In such a picture, {\it any} typical state could serve as the seed around which we build a code subspace. We could define multiple code subspaces in this way, moving their center around by picking different seed typical states.  In this way, the full space would be in some sense a product of a code subspace and something which is spanned in a redundant way by microstates themselves. Of course, this redundant formulation does not present the nice direct sum structure that we described, but perhaps there is a way of making this notion precise. 

A notable feature of black hole physics is that the dynamics amongst microstates after a perturbation appears to be chaotic, perhaps leading to scrambling of information.   How does this work in our scenario?  The BPS sector that we have investigated is integrable and does not therefore have a truly chaotic dynamics.  However, any small perturbation out of the BPS subspace will engage the full Yang-Mills theory and its chaotic dynamics.  Even more interestingly, it is possible that coarse-graining an integrable system can cause the low-energy dynamics to look thermalizing or scrambling.  The reason is that the IR is an open quantum system relative to the UV, and the UV can act effectively as a thermal bath \cite{CALDEIRA1983587}.  (See also a recent treatment in \cite{Agon:2014uxa}.)


Finally, our paper has a bearing on the question of why black hole entropy is proportional to a geometric quantity, the horizon area.
On the one hand, there is significant evidence that black holes are universal effective gravitational descriptions of a very large number of microstates, and the entropy counts this microscopic degeneracy \cite{Strominger:1996sh, Lunin:2002qf, Balasubramanian:2005mg, Balasubramanian:2007zt, Balasubramanian:2008da}. But this  degeneracy comes from a detailed ultraviolet completion of quantum gravity, so the above counting does not give a conceptual gravitational explanation for why the entropy is given by the Bekenstein-Hawking formula $S_{BH}=A/4G_N$. Indeed, this expression can be derived from purely semi-classical GR arguments without any reference to the UV. On the other hand, it has been suggested that there is an alternative description of black hole entropy in terms of counting soft modes, or ``edge modes'', at the horizon of the black hole \cite{Hawking:2016msc, Haco:2018ske, Donnay:2016ejv}. While this line of reasoning may lead to a more conceptual understanding of the Bekenstein-Hawking formula within gravity, it is unclear how it relates to the microstate counting arguments. In this paper, we found a concrete example where these two different approaches have a chance of converging. The reason we can make progress is because the bulk surface where the boundary field theory lives is identified with the stretched horizon and thus we can understand its physics directly in terms of boundary variables, using a coarse grained version of the standard LLM dictionary. This is to be contrasted with the usual $AdS/CFT$ setup where the stretched horizon and the boundary CFT are very far from each other and the mapping becomes very non-local.  On the gravity side, we identified a soft mode which becomes physical when introducing 
a stretched horizon. We showed that from the CFT point of view the stretched horizon corresponds to a coarse-graining, and that the soft mode is related via a canonical transformation to ultraviolet phase space degrees of freedom, which are essentially ``microstate deformations''. Surprisingly, the soft modes are purely IR effective modes, although they encode information about the microstructure of the black hole.   This is a specific failure of decoupling in gravity, such that the IR retains certain  information about the UV, and encodes this data in geometry.

\section*{Acknowledgements}
We thank Gautam Mandal, Gabor Sarosi, Matthew Headrick, Albion Lawrence and Bogdan Stoica for helpful conversations. VB and OP acknowledge support from the Simons Foundation (\# 385592, VB) through the It
From Qubit Simons Collaboration, and the US Department of Energy contract \# FG02-
05ER-41367. DB and AM work  supported in part by the department of Energy under grant {DE-SC} 0011702.  AL acknowledges support from the Simons Foundation through the It from
Qubit collaboration. AL would also like to thank the Department of Physics and Astronomy
at the University of Pennsylvania for hospitality during the development of this work. 
CR wishes to acknowledge support from the Simons Foundation (\#385592, VB) through the It From Qubit Simons Collaboration, from the Belgian Federal Science Policy Office through the Interuniversity Attraction Pole P7/37, by FWO-Vlaanderen through projects G020714N and G044016N, and from Vrije Universiteit Brussel through the Strategic Research Program ``High-Energy Physics''. VB, AL, OP and CR would like to thank the Galileo Galilei Institute for Theoretical Physics, Florence for their hospitality during the workshop ``Entanglement in quantum systems'', where part of this work was carried out. 

\appendix

\section{Further details on Symplectic form}\label{appA}
In the main text, we computed the gravitational symplectic form within the bulk of the grey disc in region (i), and simply stated that the contributions from the boundary of the disc, i.e., region (ii), and from outside the disc, i.e., region (iii), vanish. Here, we verify this claim. 

Let us begin with region (iii). Here it is convenient to break up $\bs{\omega}$ as 
$$\bs{\omega} = \bs{\omega}_I + \bs{\omega}_{II}.$$
where $\bs{\omega}_I$ consists of terms which do not depend on $(\lambda, \tilde{\lambda})$, and $\bs{\omega}_{II}$ consists of terms which do depend on $(\lambda, \tilde{\lambda})$. Let us first focus on the $\lambda$-independent terms. In region (iii), these reduce to 
\beqn
\bs\omega_I &=& -\epsilon_{ij}  \bd V_i \bd \left(\alpha V_j\right) - \epsilon_{ij} \bd\left[\frac{y^2 V_i}{2\left(\frac{1}{4} - z^2\right)}+ U_i\right] \bs \bd\left[\frac{y^2 z V_j}{\left(\frac{1}{4} - z^2\right)}\right]\nonumber\\
&=&\epsilon_{ij}  \frac{y^4 z\left(\frac{1}{4} + z^2\right)}{\left(\frac{1}{4} - z^2\right)^2} \bd V_i \bd  V_j - \epsilon_{ij} \frac{y^4 z}{2\left(\frac{1}{4} - z^2\right)^2}\bd V_i\bd V_j -\epsilon_{ij}\frac{y^2 z }{\left(\frac{1}{4} - z^2\right)} \bd U_i \bd V_j
\eeqn 
where in the second line we have used the fact that $\bd z_0$ vanishes in region (iii), so the $\bd$ can only act on $V_i$ or $U_i$. Now the first two terms above cancel in region (iii)\footnote{Equivalently, the coefficient of the $\bd V_i \bd V_j$ term goes as $\frac{y_0^4}{(1/4-z^2)}$ which vanishes as $y_0 \to 0$.}, so we have
\beq
\bs{\omega} = -\epsilon_{ij}\frac{y^2 z }{\left(\frac{1}{4} - z^2\right)} \bd U_i \bd V_j.
\eeq
Next, let's consider the gauge-dependent terms 
\beq
\bs\omega_{II}  = 4 \epsilon^{ij} \left(\bd \lambda \bd \tilde{F}_{ij} - \bd \tilde{\lambda} \bd F_{ij}\right).
\eeq
In region (iii), the flux corresponding to $\bd U_i$ vanishes, this becomes
\beqn
\bs\omega_{II}  &=& -8 \epsilon^{ij} \left\{\bd \lambda  \pa_i \left[\frac{y^2}{4(\frac{1}{2}+z)}\bd V_j\right] - \bd \tilde{\lambda}\pa_i \left[\frac{y^2}{4(\frac{1}{2}-z)}\bd V_j\right]\right\}\nonumber\\
&=&8 \epsilon^{ij}\bd \tilde{\lambda}\pa_i \left[\frac{y^2}{4(\frac{1}{2}-z)}\bd V_j\right]\nonumber\\
&=&-2 \epsilon^{ij}\frac{y^2}{(\frac{1}{2}-z)}\pa_i\bd \tilde{\lambda} \bd V_j+2 \pa_i \left\{\epsilon^{ij}\frac{y^2}{(\frac{1}{2}-z)}\bd \tilde{\lambda} \bd V_j \right\}
\eeqn
Importantly, in region (iii) we have the regularity condition $\pa_i\bd \tilde{\lambda} = \bd\tilde{B}_i = -\frac{1}{4}\bd U_i$. So this gives
\beq
\bs{\omega}_{II} =  \frac{1}{2}\epsilon_{ij}\frac{y^2}{(\frac{1}{2}-z)}\bd U_i\bd V_j+2 \pa_i \left\{\epsilon^{ij}\frac{y^2}{(\frac{1}{2}-z)}\bd \tilde{\lambda} \bd V_j \right\}
\eeq
Combining $\bs{\omega}_I$ and $\bs{\omega}_{II}$ we get
\beq
\bs{\Omega}_{ext.} = 2 \oint_{r=R+\varepsilon} \frac{y^2}{(\frac{1}{2}-z)}\bd \tilde{\lambda} \bd V_{\parallel} 
\eeq
where $\tilde{\lambda}$ is determined by the regularity condition on region (iii): $\pa_i\tilde{\lambda} = \frac{1}{4}U_i$. Note that the factor $\frac{y^2}{(\frac{1}{2}-z)}$ vanishes at $r=R$\footnote{This follows from writing $z = \frac{1}{2} + y_0^2 f(r)+\cdots,$ and checking that $\lim_{r\to R} f(r) = \infty.$} (i.e. the order of limits is we send $y \to 0$ first and then we send $\varepsilon \to 0$), but we must be careful because $\bd V_{\parallel}$ might diverge in this limit. However, as long as we consider greyscale deformations (i.e. $\bd z_{0}$ is localized far from the boundary) then this does not happen. So in this case, we conclude that the symplectic form from outside indeed vanishes.  

Since we do not expect any singular behavior in the symplectic form from the boundary region for greyscale deformations, the contribution from region (ii) similarly vanishes as we send $ \varepsilon \to 0$.

\section{Symplectic form for large deformations}\label{applargedef}
In the section \ref{SFCFT}, we constructed the coarse-grained symplectic form from the CFT, and showed the emergence of a new effective mode in the IR built out of the UV modes in the phase space density.  In this analysis we considered localized perturbations of phase that correspond on the AdS side to graviton fluctuations.   In this appendix, we consider large deformations, e.g. changing the radii or widths of the rings in phase space.  In gravity these are large perturbations that globally move the locations and sizes of  spheres in the geometry.  After coarse-graining a typical state, such perturbations will appear as angularly-symmetric greyscale deformations (as opposed local perturbations).   


To study such large deformations, let us begin with the observation that the slope $y'(x)$ of the typical Young tableau at some point $x$ is determined by $\langle c_{N-x}\rangle$, where $c_i$ is the number of columns of length $i$.  Here $x$ is the row-number in the tableau, and should be confused with $x$  in the previous section, where it was a coordinate on phase space.  So, the angularly-symmetric greyscale deformations are directly related to the change in the one-point functions of $c_k$s. Let us review this relation first. Our starting point is a formula for the greyscale factor $u = \frac{1}{2}-z_0$ in terms of the shape of the typical Young tableau in the CFT description (\cite{Balasubramanian:2005mg}):
\beq
u(0,r^2) = \frac{1}{1+y'(x)}, 
\eeq
where on the right hand side we should think of $x$ as a function of $r$ obtained implicitly from the relation\footnote{This relation basically says that $y(x) + x$ is the energy associated the fermion at row-position number $x$.}
\beq
y(x) + x = \frac{r^2}{2\hbar}.
\eeq 
This formula directly relates greyscale deformations to shape-deformations of the typical Young tableau. 
\begin{figure}[!h]
\centering
\includegraphics[height=3cm]{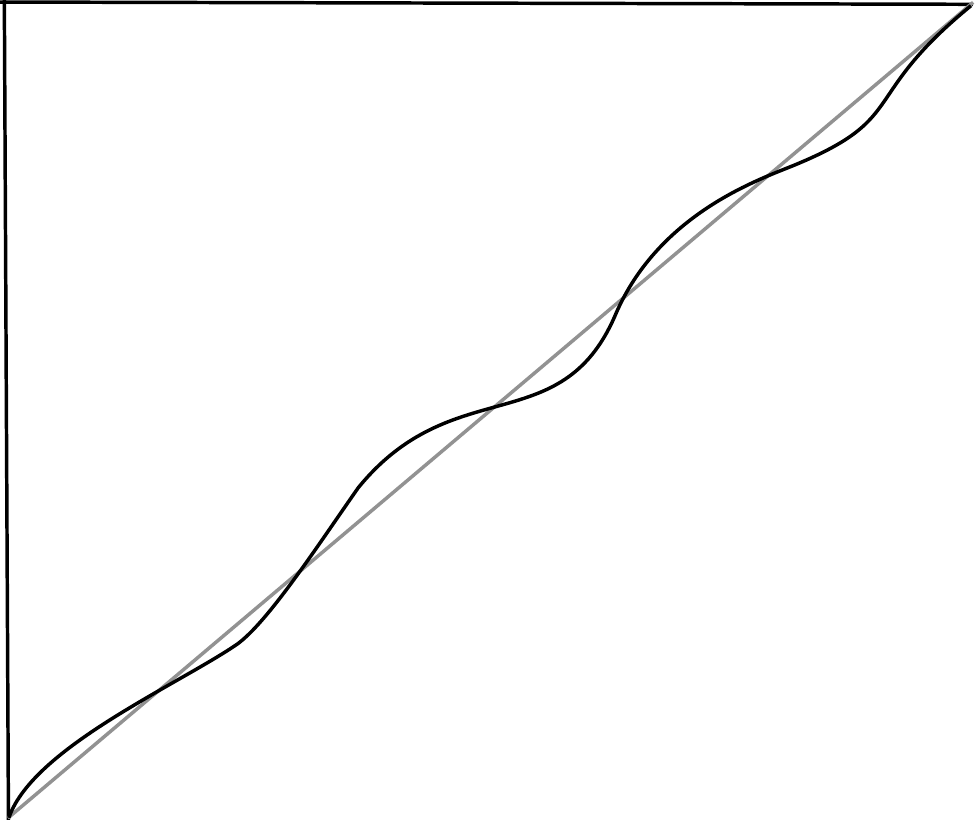}
\caption{\textsf{\small{A shape-deformation of the triangular Young tableau.}}}
\end{figure}

Now consider a shape deformation of the tableau $y(x) = y_0(x)+\delta y(x)$, where $\delta y$ is an infinitesimal ripple of the original tableau (see figure). We will mostly work to linear order in $\delta y$. It is a simple matter to compute the change in the phase space denisty $u$. Let $x_0$ be the solution to the equation
\beq
y(x_0) +x_0 = \frac{r^2}{2\hbar}.
\eeq
Then under a change in $y$, the change in the solution $\delta x(r^2) $ is given by
\beq
\delta x = -\frac{\delta y(x_0)}{1+ y'(x_0)}.
\eeq
Using this to compute $\delta u$, we obtain
\beq
\delta u(r^2) = -  \frac{1}{(1+y'(x_0))^2}\left(\delta y'(x_0) - \frac{\delta y(x_0)}{1+y'(x_0)}y''(x_0)\right).
\eeq
We can now evaluate this around the superstar $y_0(x) = \omega x$ where $\omega = \frac{N_c}{N}$, which gives $x_0(r^2) = \frac{r^2}{2(1+\omega)\hbar}$. So we conclude that
\beq \label{du}
\delta u(r^2)  = -\frac{1}{(1+\omega)^2}\delta y'\left(x_0\right)= -\frac{1}{4} \delta \langle c_{N-x_0} \rangle,
\eeq
where in the last equality we are considering the $\omega = 1$ superstar.  This gives a direct relation between angularly-symmetric grey-scale deformations of the LLM geometry and one-point functions of $c_k$s. 


Let us now go back to the superstar, and in particular to the angularly symmetric mode $\delta u=  - \frac{1}{4} \delta \langle c \rangle$. Let $(D_k, D_k^{\dagger})$ (for $k=1,2\cdots N$) be operators which annihilate and create columns of length $k$ \cite{Dhar:2005fg}. In other words, 
\begin{eqnarray}
D_k |c_1,\cdots ,c_k,\cdots c_N\rangle &=& \sqrt{c_k} |c_1,\cdots ,c_k-1,\cdots c_N\rangle, \\
D^{\dagger}_k |c_1,\cdots ,c_k,\cdots c_N\rangle &=& \sqrt{c_k+1} |c_1,\cdots ,c_k+1,\cdots c_N\rangle. 
\end{eqnarray}
By definition, these operators satisfy
\beq
\left[ D_k, D^{\dagger}_m \right] = \delta_{km}, \;\; c_k = D_k^{\dagger}D_k.
\eeq

A symplectic form is only defined for classical fluctuations on phase space, which we can extract by constructing a class of ``coherent states'' where the number operator $c_k$ and the phase operator are both classical. To begin with, we want construct the superstar background as a coherent state above the vacuum.  To do this consider the state
\beq
|\{z_k\} \rangle = \exp\left( \sum_k z_k D^{\dagger}_k - z^*_k D_k \right) |0,\cdots,0\rangle.
\eeq
This state is of course classical with respect to $D_k$ and $D^{\dagger}_k$. So, we can associate a well-defined symplectic form to fluctuations in these variables:
\beq
\bs{\Omega} = \sum_k \delta z_k \wedge \delta z_k^*.
\eeq
The $\delta z$ should coarse-grain to be the angularly symmetric greyscale fluctuations around the superstar.

Now, it is tempting to simply change coordinates to $z_k = \sqrt{u_k} e^{i\pi_k}$, and rewrite the symplectic form in terms of the density and phase, which would be canonically conjugate.  The density is of course related to the number operator $c_k$. However, there is a subtlety. In order for the background to be close to the superstar, we must take $|z_k|=1,\;\;\forall k$ so as to have $\langle c_k \rangle = 1$.  However, because the variance of the number operator in a coherent state satisfies 
\beq
\frac{\Delta c_k}{\langle c_k \rangle } = \frac{1}{|z_k|} \, ,
\eeq
we see that for the superstar, which has $|z_k| \sim 1$ for all $k$, the variance is comparable to the mean, meaning that the state is \emph{not} classical with respect to the number operator.
Thus if we write $z_k = \sqrt{u_k} e^{i\pi_k}$ then $u_k$ is not a classical variable even though $z_k$ is. In the spirit of section $5$, a natural resolution is to consider long wavelength modes for the $z_k$: that is we don't have access to the individual $k$ labels but to long wavelengths superposition of these. In this case,  we expect the variance of the long wavelength $c$ to be $1/N$ suppressed with respect to the mean. 

\begin{figure}[!h]
\centering
\includegraphics[height=4cm]{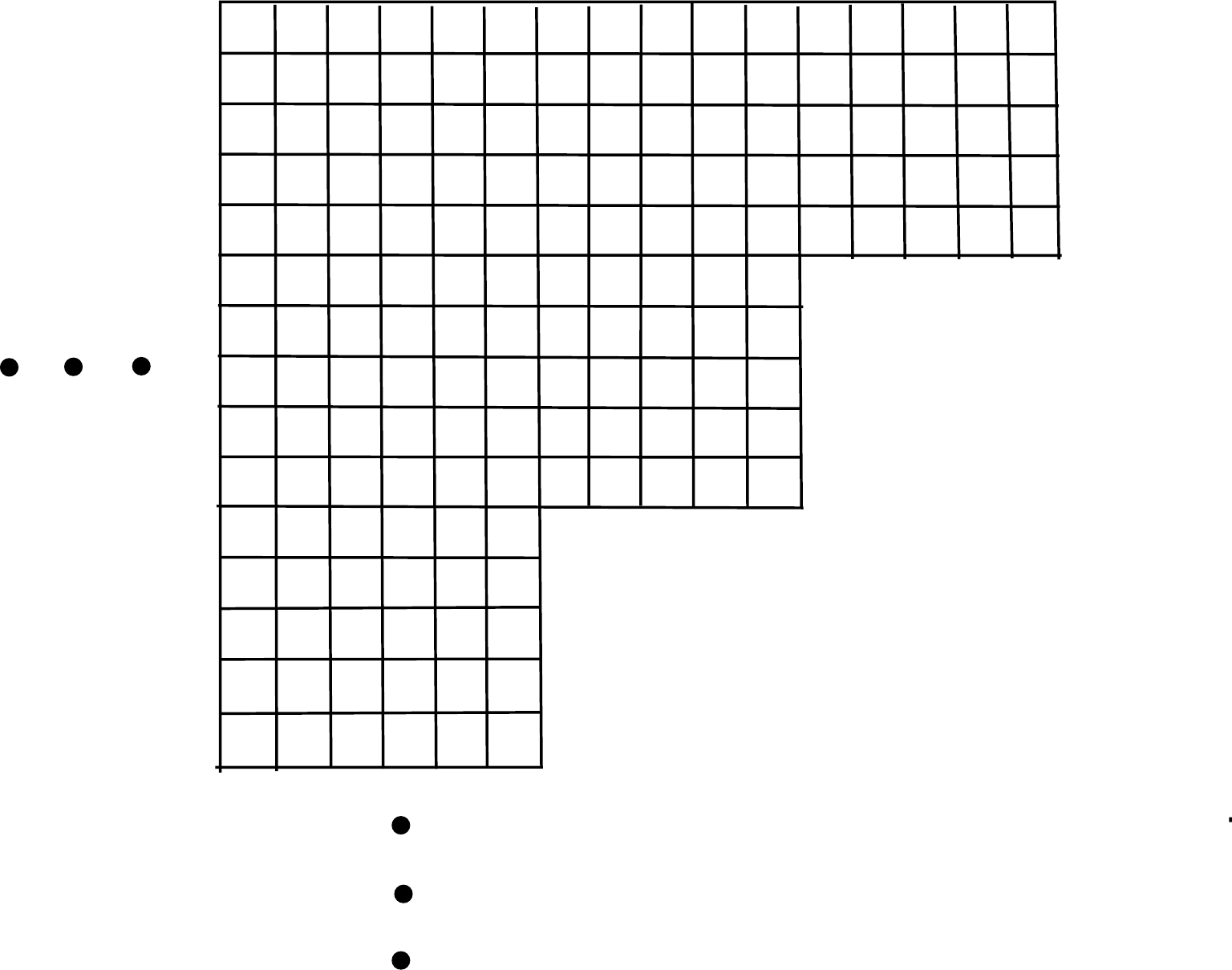}
\caption{\small{\textsf{The coherent states which we consider resemble the stair-case tableaus, but have the advantage that they are classical with respect to the number operator and the phase operator.\label{staircase}}}}
\end{figure}

However, if $|z_k|$ becomes large-but-not-too-large (scaling with say $N^{1/4}$) for some $k$, then the ratio of the variance to the one-point function of the number operator becomes small, and hence the number operator becomes classical, without need for further coarse graining. Then, since $z_k$ and $z_k^*$ are already classical in any coherent states, the phase must also be classical.  States with these properties look like \emph{staircases} in the Young tableau description (Fig.~\ref{staircase}).  These states have microscopic structure that is sufficiently far from the $\hbar$ scale for us to be able to do a phase space analysis.


Following Sec.~\ref{SFGrav} the staircase states correspond in gravity to a concentric black and white ring boundary condition in the LLM plane.  Deformations that increase or decrease the number of columns of a given length correspond in gravity to increasing or decreasing the widths of some rings. For such deformations we can write a symplectic form as
\beq
\bs{\Omega} = \sum_n \delta u_{n} \wedge  \delta \pi_n,
\eeq
where $n$ indexes column lengths in the reference tableau (e.g. Fig.~\ref{staircase}).  This construction is inherently coarse-grained because the reference tableau is only permitted to have steps that scale as $N^{1/4}$. However, we may be able to further refine this by considering a coarse-graining over the $n$-index. In this context, the density and phase are realized as classically conjugate variables. Here, once again we see that the conjugate momentum to $u$ is a new emergent field $\pi$, which in the present case corresponds to the phase with respect to $D_k$ and $D_k^{\dagger}$.  In order to measure the phase in some state, we need to have access to these $(D_k, D_k^{\dagger})$, which are heavy operators, and so in this sense $\pi$ is an effective IR mode which is built out of UV degrees of freedom.

\providecommand{\href}[2]{#2}\begingroup\raggedright\endgroup

\end{document}